\documentclass[ALICE,manyauthors]{cernphprep}
\usepackage[comma,square,numbers,sort&compress]{natbib}

\usepackage{color}
\usepackage{amsmath}
\usepackage{amssymb}
\usepackage{lineno}
\modulolinenumbers[5]
\usepackage{xspace}
\usepackage{hyperref}
\usepackage{makecell}
\usepackage{multirow, array}
\usepackage[thinlines]{easytable}
\usepackage[T1]{fontenc}
\usepackage{orcidlink}

\begin{document}
%

\newcommand{\pp}           {pp\xspace}
\newcommand{\ppbar}        {\mbox{$\mathrm {p\overline{p}}$}\xspace}
\newcommand{\ep}           {ep\xspace}
\newcommand{\XeXe}         {\mbox{Xe--Xe}\xspace}
\newcommand{\PbPb}         {\mbox{Pb--Pb}\xspace}
\newcommand{\pA}           {\mbox{pA}\xspace}
\newcommand{\pPb}          {\mbox{p--Pb}\xspace}
\newcommand{\AuAu}         {\mbox{Au--Au}\xspace}
\newcommand{\dAu}          {\mbox{d--Au}\xspace}

\newcommand{\s}            {\ensuremath{\sqrt{s}}\xspace}
\newcommand{\snn}          {\ensuremath{\sqrt{s_{\mathrm{NN}}}}\xspace}
\newcommand{\pt}           {\ensuremath{p_{\rm T}}\xspace}
\newcommand{\meanpt}       {$\langle p_{\mathrm{T}}\rangle$\xspace}
\newcommand{\ycms}         {\ensuremath{y_{\rm CMS}}\xspace}
\newcommand{\ylab}         {\ensuremath{y_{\rm lab}}\xspace}
\newcommand{\etarange}[1]  {\mbox{$\left | \eta \right |~<~#1$}}
\newcommand{\yrange}[1]    {\mbox{$\left | y \right |~<~#1$}}
\newcommand{\dndy}         {\ensuremath{\mathrm{d}N_\mathrm{ch}/\mathrm{d}y}\xspace}
\newcommand{\dndeta}       {\ensuremath{\mathrm{d}N_\mathrm{ch}/\mathrm{d}\eta}\xspace}
\newcommand{\avdndeta}     {\ensuremath{\langle\dndeta\rangle}\xspace}
\newcommand{\dNdy}         {\ensuremath{\mathrm{d}N_\mathrm{ch}/\mathrm{d}y}\xspace}
\newcommand{\Npart}        {\ensuremath{N_\mathrm{part}}\xspace}
\newcommand{\Ncoll}        {\ensuremath{N_\mathrm{coll}}\xspace}
\newcommand{\dEdx}         {\ensuremath{\textrm{d}E/\textrm{d}x}\xspace}
\newcommand{\RpPb}         {\ensuremath{R_{\rm pPb}}\xspace}

\newcommand{\nineH}        {$\sqrt{s}~=~0.9$~Te\kern-.1emV\xspace}
\newcommand{\seven}        {$\sqrt{s}~=~7$~Te\kern-.1emV\xspace}
\newcommand{\thirteen}     {$\sqrt{s}~=~13$~Te\kern-.1emV\xspace}
\newcommand{\twoH}         {$\sqrt{s}~=~0.2$~Te\kern-.1emV\xspace}
\newcommand{\twosevensix}  {$\sqrt{s}~=~2.76$~Te\kern-.1emV\xspace}
\newcommand{\five}         {$\sqrt{s}~=~5.02$~Te\kern-.1emV\xspace}
\newcommand{\onenn}         {$\sqrt{s}~=~1.96$~Te\kern-.1emV\xspace}

\newcommand{\twosevensixnn}{$\sqrt{s_{\mathrm{NN}}}~=~2.76$~Te\kern-.1emV\xspace}
\newcommand{\fivenn}       {$\sqrt{s_{\mathrm{NN}}}~=~5.02$~Te\kern-.1emV\xspace}
\newcommand{\twohundred}       {$\sqrt{s_{\mathrm{NN}}}~=~200$~Ge\kern-.1emV\xspace}
\newcommand{\eightsixteen} {$\sqrt{s_{\mathrm{NN}}}~=~8.16$~Te\kern-.1emV\xspace}
\newcommand{\LT}           {L{\'e}vy-Tsallis\xspace}
\newcommand{\GeVc}         {Ge\kern-.1emV/$c$\xspace}
\newcommand{\MeVc}         {Me\kern-.1emV/$c$\xspace}
\newcommand{\TeV}          {Te\kern-.1emV\xspace}
\newcommand{\GeV}          {Ge\kern-.1emV\xspace}
\newcommand{\GeVsq}          {Ge\kern-.1emV\ensuremath{^2}\xspace}
\newcommand{\MeV}          {Me\kern-.1emV\xspace}
\newcommand{\GeVmass}      {Ge\kern-.1emV/$c^2$\xspace}
\newcommand{\MeVmass}      {Me\kern-.1emV/$c^2$\xspace}
\newcommand{\lumi}         {\ensuremath{\mathcal{L}}\xspace}

\newcommand{\ITS}          {\rm{ITS}\xspace}
\newcommand{\TOF}          {\rm{TOF}\xspace}
\newcommand{\ZDC}          {\rm{ZDC}\xspace}
\newcommand{\ZDCs}         {\rm{ZDCs}\xspace}
\newcommand{\ZNA}          {\rm{ZNA}\xspace}
\newcommand{\ZNC}          {\rm{ZNC}\xspace}
\newcommand{\SPD}          {\rm{SPD}\xspace}
\newcommand{\SDD}          {\rm{SDD}\xspace}
\newcommand{\SSD}          {\rm{SSD}\xspace}
\newcommand{\TPC}          {\rm{TPC}\xspace}
\newcommand{\TRD}          {\rm{TRD}\xspace}
\newcommand{\VZERO}        {\rm{V0}\xspace}
\newcommand{\VZEROA}       {\rm{V0A}\xspace}
\newcommand{\VZEROC}       {\rm{V0C}\xspace}
\newcommand{\Vdecay} 	   {\ensuremath{V^{0}}\xspace}
\newcommand{\AD}           {\rm{AD}\xspace}
\newcommand{\ADA}          {\rm{ADA}\xspace}
\newcommand{\ADC}          {\rm{ADC}\xspace}

\newcommand{\ee}           {\ensuremath{e^{+}e^{-}}} 
\newcommand{\pip}          {\ensuremath{\pi^{+}}\xspace}
\newcommand{\pim}          {\ensuremath{\pi^{-}}\xspace}
\newcommand{\kap}          {\ensuremath{\rm{K}^{+}}\xspace}
\newcommand{\kam}          {\ensuremath{\rm{K}^{-}}\xspace}
\newcommand{\pbar}         {\ensuremath{\rm\overline{p}}\xspace}
\newcommand{\kzero}        {\ensuremath{{\rm K}^{0}_{\rm{S}}}\xspace}
\newcommand{\lmb}          {\ensuremath{\Lambda}\xspace}
\newcommand{\almb}         {\ensuremath{\overline{\Lambda}}\xspace}
\newcommand{\Om}           {\ensuremath{\Omega^-}\xspace}
\newcommand{\Mo}           {\ensuremath{\overline{\Omega}^+}\xspace}
\newcommand{\X}            {\ensuremath{\Xi^-}\xspace}
\newcommand{\Ix}           {\ensuremath{\overline{\Xi}^+}\xspace}
\newcommand{\Xis}          {\ensuremath{\Xi^{\pm}}\xspace}
\newcommand{\Oms}          {\ensuremath{\Omega^{\pm}}\xspace}
\newcommand{\degree}       {\ensuremath{^{\rm o}}\xspace}

\newcommand{\jpsi}         {\ensuremath{{\mathrm J}/\psi}\xspace}
\newcommand{\psip}         {\ensuremath{\psi(2\mathrm{S})}\xspace}
\newcommand{\fD}           {\ensuremath{f_{\rm D}}\xspace}
\newcommand{\AxE}          {\ensuremath{A\times{\epsilon}}}
\newcommand{\starlight}    {STARlight\xspace}
\newcommand{\superchic}    {SuperChic\xspace}
\newcommand{\ggmm}         {\ensuremath{\gamma \gamma \rightarrow \mu^+ \mu^-}\xspace}
\newcommand{\gp}         {\ensuremath{\gamma \, \mathrm{p}}\xspace}
\newcommand{\gPb}         {\ensuremath{\gamma \, \mathrm{Pb}}\xspace}
\newcommand{\Wgp}          {\ensuremath{W_{\gp}}\xspace}
\newcommand{\mmm}         {\ensuremath{M_{\mu\mu}}\xspace}

\begin{titlepage}
\PHyear{2023}       
\PHnumber{059}      
\PHdate{29 March}   

\title{Exclusive and dissociative \jpsi photoproduction, and exclusive dimuon production, in \pPb collisions at \eightsixteen}

\ShortTitle{Dimuon continuum, and exclusive and dissociative \jpsi production in UPCs}

\Collaboration{ALICE Collaboration\thanks{See Appendix~\ref{app:collab} for the list of collaboration members}}
\ShortAuthor{ALICE Collaboration} 

\begin{abstract}

The ALICE Collaboration reports three measurements in ultra-peripheral proton--lead collisions at forward rapidity.  The exclusive two-photon process \ggmm and the exclusive photoproduction of \jpsi are studied. \jpsi photoproduction with proton dissociation is measured for the first time at a hadron collider. 
The cross section for the two-photon process of dimuons in the invariant mass range from 1 to 2.5 GeV/$c^2$ agrees with leading order quantum electrodynamics calculations. 
The exclusive and dissociative cross sections for \jpsi photoproductions are measured for photon--proton centre-of-mass energies from 27 to 57 GeV. They are 
in good agreement with HERA results.

\end{abstract}
\end{titlepage}

\setcounter{page}{2} 


\section{Introduction}

The strong electromagnetic fields present in ultra-peripheral collisions (UPCs) offer a unique opportunity to study a variety of phenomena, such as photonuclear and two-photon processes~\cite{Baltz:2007kq, contreras2015ultra, Klein:2020fmr}. 
These interactions are mediated by quasireal photons and characterised by an impact parameter larger than the sum of the radii of  the colliding nuclei.

Two-photon interactions can give rise to exclusive non-resonant dimuon production.
Precise measurements of this process 
can be used to test quantum electrodynamics (QED) calculations, 
such as light-by-light scattering~\cite{dEnterria:2013zqi, Klusek-Gawenda:2016euz} recently measured by ATLAS~\cite{PhysRevLett.123.052001, ATLAS:2020hii, ATLAS:2017fur} and CMS~\cite{2019134826}, and higher-order QED effects~\cite{Zha:2021jhf}. The latter are expected to be sizeable,  since the photon couples to nuclei with a large coupling $Z\alpha$ where $Z$ is the charge number and $\alpha$ is the fine structure constant~\cite{Zha:2021jhf}. 
Various theoretical calculations predict a different strength of higher-order effects in heavy-ion collisions~\cite{Hencken:2006ir,Baltz:2010mc,Zha:2021jhf}. The use of asymmetric p–Pb collisions may provide additional insight on higher-order corrections from multi-photon exchange with a single ion~\cite{Baltz:2007kq}.

Measurements of cross sections of dilepton production using UPC samples, performed by ALICE in \PbPb~\cite{ALICE:2013wjo} and \pPb~\cite{ALICE:2018oyo}, CMS~\cite{CMS:2020skx} and ATLAS~\cite{ATLAS:2022vbe} in  \PbPb, and PHENIX~\cite{Afanasiev:2009hy} and STAR~\cite{STAR:2004bzo, STAR:2019wlg, STAR:2018ldd} in \AuAu, are consistent with leading-order (LO) QED calculations. 
However, latest precision measurements by ATLAS~\cite{ATLAS:2020epq} revealed a significant discrepancy with LO QED predictions from the \starlight event generator~\cite{Klein:2016yzr}, up to 20\% at large rapidities. This discrepancy is discussed by the authors of the \superchic event generator in Ref.~\cite{Harland-Lang:2021ysd}. They argue that \starlight does not take into account contributions from dilepton--nucleus impact parameters smaller than the nuclear radii, expected to be significant at high photon energies. Accounting for photons emitted at such impact parameters is also discussed in Refs.~\cite{Burmasov:2021phy,Zha:2021jhf,Azevedo:2019fyz}.
However, this effect alone does not allow the \superchic authors to resolve the discrepancy with the ATLAS data~\cite{Burmasov:2021phy}. 
The inclusion of higher-order corrections to the LO QED calculation could explain the ATLAS results, as suggested in Ref.~\cite{Zha:2021jhf}. 
Dilepton measurements were also performed in \pp collisions by ATLAS~\cite{ATLAS:2015wnx, ATLAS:2017sfe} and CMS~\cite{CMS:2011vma}, and in  \ppbar collisions by CDF~\cite{CDF:2006apx,CDF:2009xey}.
These measurements did not explore  the low invariant mass region at forward rapidities.

Dimuons can also be produced in photonuclear reactions, from the decay of a vector meson. In particular, dimuons can be produced from the decay of a \jpsi meson in the elastic process $\gamma + {\rm p} \rightarrow \jpsi + {\rm p}$, or with proton dissociation in the reaction $\gamma + {\rm p} \rightarrow \jpsi + {\rm p}^{(*)}$. 
The use of \pPb collisions offers the possibility of assigning the photon to its source: the lead ion is in most of the cases the photon emitter due to its large charge number.  
The \gp centre-of-mass energy $\Wgp$ is a function of the \jpsi rapidity:
$\Wgp^2 = 2 E_{\rm p} M_{\jpsi} \exp(-y)$, where $M_{\jpsi}$ is the \jpsi mass, $y$ is the \jpsi rapidity measured in the laboratory frame with respect to the proton beam direction, and $E_{\rm p} = 6.5$~\TeV is the proton beam energy, corresponding to a centre-of-mass energy in the \pPb system of \eightsixteen. The energy range studied is $27 < \Wgp < 57$~\GeV, which corresponds to a longitudinal momentum fraction of the participating partons, Bjorken-$x$ scale, in the range $5 \times 10^{-3} < x < 2 \times 10^{-2}$, where the conversion is performed as $x = (M_{\jpsi}/\Wgp)^2$. This is a similar kinematic domain as studied at HERA~\cite{Newman:2013ada}.

Exclusive \jpsi photoproduction is sensitive to the gluon distribution in protons, since its cross section scales with the square of the gluon parton density function in the target proton, according to LO QCD calculations~\cite{Ryskin:1992ui}. This picture may change at next-to-leading order (NLO) according to the recent studies in Ref.~\cite{Eskola:2022vpi}. At high \Wgp, a reduction in the growth rate of the exclusive \jpsi photoproduction cross section would indicate that non-linear 
 QCD dynamics are present. 
These non-linearities may arise from gluon recombination, which tame the growth of the gluon distribution, leading in the high energy limit to the gluon saturation phenomenon~\cite{Morreale:2021pnn}.

On the other hand, \jpsi photoproduction off protons with proton dissociation is a scattering event that produces a \jpsi vector meson and, accompanied by a large rapidity gap, remnants of the dissociated proton.
This process might serve as an experimental signature of 
subnucleonic fluctuations of initial state configurations in the  proton target~\cite{Miettinen:1978jb,Mantysaari:2020axf, Cepila:2016uku}. At high energies, the ratio of dissociative to exclusive cross sections is predicted to vanish, owing to the onset of gluon saturation at sufficiently small $x$~\cite{Cepila:2016uku,Mantysaari:2020axf}.

At HERA, ZEUS and H1 have measured both the exclusive~\cite{ZEUS:2002wfj, H1:2005dtp, H1:2013okq} and dissociative~\cite{ZEUS:2009ixm, H1:2013okq} \jpsi photoproduction off protons at \gp centre-of-mass energies ranging from 20 to 305~\GeV. 
CDF has measured the exclusive process in \ppbar collisions at \onenn~\cite{CDF:2009xey}.
At the LHC, the exclusive process was studied in \pPb at \fivenn by ALICE~\cite{ALICE:2014eof,ALICE:2018oyo}, and in \pp at \seven and \thirteen by LHCb~\cite{LHCb:2013nqs, Aaij:2014iea, LHCb:2018rcm}. 
The dissociative process has never been measured before at a hadron collider.

In this article, the measurement of exclusive dimuon continuum production in two-photon interactions in \pPb UPCs at \eightsixteen is presented. It is performed in three intervals of dimuon invariant mass, in the range $1.0 < \mmm < 2.5$~\GeVmass, and two intervals of rapidity, in the range $2.5 < y < 4.0$.
The measurement of exclusive \jpsi photoproduction off protons is also presented along with 
the measurement of \jpsi photoproduction with proton dissociation.  These three measurements are carried out in the forward rapidity region with respect to the proton beam direction, namely  $2.5 < y < 4.0$, and at low dimuon transverse momentum, $\pt < 3$~\GeVc. This corresponds to a range in the square of the momentum transferred at the proton vertex $|t| \lesssim 9$~\GeVsq, where $t \approx -\pt^2$.

\section{Experimental setup and trigger}

The ALICE detector is described in Ref.~\cite{ALICE:2008ngc} and its performance is detailed in Ref.~\cite{Abelev:2014ffa}. 
The main ALICE tracking detector used in this analysis is the single-arm muon spectrometer, covering the pseudorapidity interval $-4.0 < \eta < -2.5$. \footnote{
In the ALICE convention, the muon spectrometer lies at negative longitudinal coordinate $z$, where $z = 0$ is the nominal interaction point position. therefore at negative pseudorapidity.
}. 
The analysis also uses other detector systems, namely the Silicon Pixel Detector (\SPD), VZERO (\VZERO), Zero Degree Calorimeters (\ZDCs) and ALICE Diffractive (\AD) detectors.

The muon spectrometer consists of a ten hadronic interaction length absorber, followed by five tracking stations, each made of two planes of cathode pad chambers. The third station is placed inside a dipole magnet with a 3 T$\times$m integrated magnetic field. 
Muon tracks are reconstructed by the tracking algorithm described in Ref.~\cite{ALICE:2011zqe} using the five tracking stations. 
The muon trigger system, downstream of the tracking chambers, consists of four planes of resistive plate chambers placed behind a 7.2 interaction length iron wall. 
The muon tracks detected in these planes are used in the trigger and matched offline to the muon tracks reconstructed in the five tracking stations.

The central region $| \eta | < 1.4$ is covered by the \SPD consisting of two cylindrical layers of silicon pixels, from which tracklets are reconstructed.
Tracklets are track fragments created from the primary vertex and two reconstructed points in the \SPD, one in each layer.

The \VZERO detector is composed of two arrays of scintillator counters, namely the \VZEROC and \VZEROA detectors. Each array consists of 32 cells forming four concentric rings with eight sectors each. \VZEROC, placed at the longitudinal coordinate $z = -90$~cm, covers the interval $-3.7 < \eta < -1.7$, while \VZEROA, $z = 330$~cm, covers the pseudorapidity interval $2.8 < \eta < 5.1$.  
The \AD detector~\cite{LHCForwardPhysicsWorkingGroup:2016ote, Broz:2020ejr} is composed of two scintillator tile arrays, the \ADC and \ADA subdetectors, located at $z = -19.5$~m and $z = +16.9$~m and covering the pseudorapidity ranges $-7.0 < \eta < -4.9$ and $4.7 < \eta < 6.3$, respectively. 
The time resolution of \VZERO and \AD detectors is better than 1~ns, which makes it possible to discriminate between beam--beam and beam--gas events, in which beam particles interact with residual gas inside the beam pipe. 
The raw signals of the \VZERO and \AD detectors are used in the trigger.
Offline, these detectors are used to differentiate beam--beam and beam--gas interactions.

The two \ZDCs are located at $112.5$~m from the nominal interaction point along the beam axis on either side of the ALICE detector. They are used to detect neutrons emitted in the very forward region and measure timing information of signals, thus making possible the discrimination of background signals such as beam--satellite events described in Ref.~\cite{Alici:1427728}.

\label{sec:trigger}

Exclusive dimuon production from the decay of a \jpsi or from two-photon interactions has a clear experi- mental signature: the $\mu^+\mu^-$ pair in an otherwise empty detector. On the other hand, the study of \jpsi photoproduction with a dissociative proton implies that the detector might not be empty on the proton side. 
The trigger used in these analyses is required to have at least one track with a low transverse momentum threshold (\pt $\sim 0.5$~\GeVc) in the muon spectrometer trigger system, and vetoes on \VZEROA and \ADA which are located in the flight direction of the outgoing Pb ion.

The measurements presented here use a sample of events collected during the 2016 \pPb data taking period, at \eightsixteen, corresponding to an integrated luminosity of  $\mathcal{L} = 7.90 \pm 0.14$ nb$^{-1}$~\cite{ALICE-PUBLIC-2018-002}. In these collisions the incoming proton beam travelled towards the muon spectrometer.

\section{Data sample}

\subsection{Event selection}
\label{sec:event}

Besides the trigger selection, events have to fulfill additional criteria.
 First, there must be exactly two tracks with opposite electric charge reconstructed in the muon spectrometer. Both tracks are required to match muon trigger tracks with a \pt threshold above 0.5~\GeVc.
Each track pseudorapidity is required to be within the acceptance of the muon spectrometer $-4.0 < \eta < -2.5$.
To reject tracks crossing the high-density section of the front absorber, where multiple scattering and energy loss effects are large, the muon tracks are required to exit  the front absorber  at a radial distance from the beam axis $17.6 < R_{\rm abs} < 89.5$ cm. The product of the total track momentum $p$ and the distance of closest approach (DCA), defined as the distance in the transverse plane between the extrapolated position of the reconstructed track in the tracking stations and the position of the nominal interaction point, is required to be smaller than 6 times the standard deviation of the dispersion due to multiple scattering and detector resolution. This ensures that the selected muons come from the interaction vertex without rejecting signal events.
The dimuon rapidity 
has to be in the range $2.5 < y < 4.0$,
and the dimuon \pt must be less than $3$~\GeVc.

To ensure 
that the Pb ion remains intact,
the \VZEROA and \ADA are required to have no signal at the offline level. 
The neutron \ZDC on the Pb side (\ZNA) must have no activity within $\pm 2$~ns of the expected time of the collision.
In order to suppress hadronic interactions producing particles at midrapidity, events with more than two tracklets in the \SPD layers are rejected.

Finally, the number of cells with a signal over threshold in \VZEROC must be smaller than or equal to the sum of the number of fired V0C cells matched to a muon and two additional fired cells. The matching of a muon to a fired \VZEROC cell is performed by using the ($\eta, \varphi$) coordinates of each track, where $\varphi$ is the azimuthal coordinate. 
Studies with the RAPGAP 3.3 event generator~\cite{Jung:1993gf}, a Monte Carlo program used to simulate dissociative \jpsi photoproduction in electron--proton collisions, show that the proton remnants do not leave a signal in the acceptance of the \VZEROC detector. 
The requirement on the number of fired V0C cells prevents contamination from hadronic interactions at forward rapidity.

Allowing two midrapidity tracklets in the \SPD layers and two additional fired \VZEROC cells prevents detectors from vetoing events of interest due to an additional activity, such as muon bremsstrahlung or pile-up events. Pile-up events are induced mainly by independent hadronic or electromagnetic processes, e.g. dielectron production in the
$\gamma \gamma \to {\rm e}^+ {\rm e}^-$ process, accompanying the process of interest.

\subsection{Event selection with exclusivity in the proton side}
\label{sec:p-exc}

The exclusive-dominated sample is obtained by applying the following additional criteria on the proton side. The \ADC is required to 
have no signal and the neutron \ZDC on the proton side must 
have no activity within $\pm 6$~ns of the expected time of the collision. This selection is more restrictive than on the Pb side, due to an observed asymmetry of time distributions between both sides.
Furthermore, since exclusive events are expected to be dominant at low \pt, dimuons are required to have $\pt < 1.2$~\GeVc.

\section{Monte Carlo samples}
\label{sec:MC-data}

The \starlight 2.2.0 Monte Carlo generator~\cite{Klein:1999qj, Klein:2016yzr} is used to generate the following processes: 
exclusive \jpsi 
production in \gp interactions, production of $\jpsi$ in \gPb interactions, production of \jpsi events from decays of \psip in \gp interactions, and exclusive dimuon continuum production.
The decay muons are propagated through a model of the apparatus implemented in GEANT 3.21~\cite{Brun:1994aa}, and events pass through a simulation of the detector matching the data taking conditions. 
For exclusive \jpsi production, the $t$ distribution is modelled in \starlight by a function of the form $\exp(-bt)$, where $b$ is set to 3.75~GeV$^{-2}$ to better describe the \jpsi \pt distribution in data.

\section{Data analysis}

\subsection{Signal extraction for the two-photon process at low masses}
\label{sec:gg-sig}

The yields of dimuons from exclusive two-photon interactions, $N_{\gamma \gamma}$, are measured by performing an unbinned log-likelihood fit of the \pt distribution up to $\pt = 3$~\GeVc  of the selected dimuons in the invariant mass range $1.0 < \mmm  < 2.5$~\GeVmass, where no contamination is expected from the \jpsi peak. 
The measurements are performed as a function of the dimuon invariant mass, in the three intervals $1.0 < \mmm < 1.5$~\GeVmass, $1.5 < \mmm  < 2.0$~\GeVmass, and $2.0 < \mmm < 2.5$~\GeVmass. They are presented in the rapidity interval $ 2.5 < y < 4.0$, and for $ 2.5 < y < 3.25$ and $3.25 < y < 4.0$, where the rapidity is measured in the laboratory frame with respect to the proton beam direction. 

\begin{figure}[tb]
    \begin{center}
 \includegraphics[width=0.58\textwidth, trim={0 0 0 0},clip]{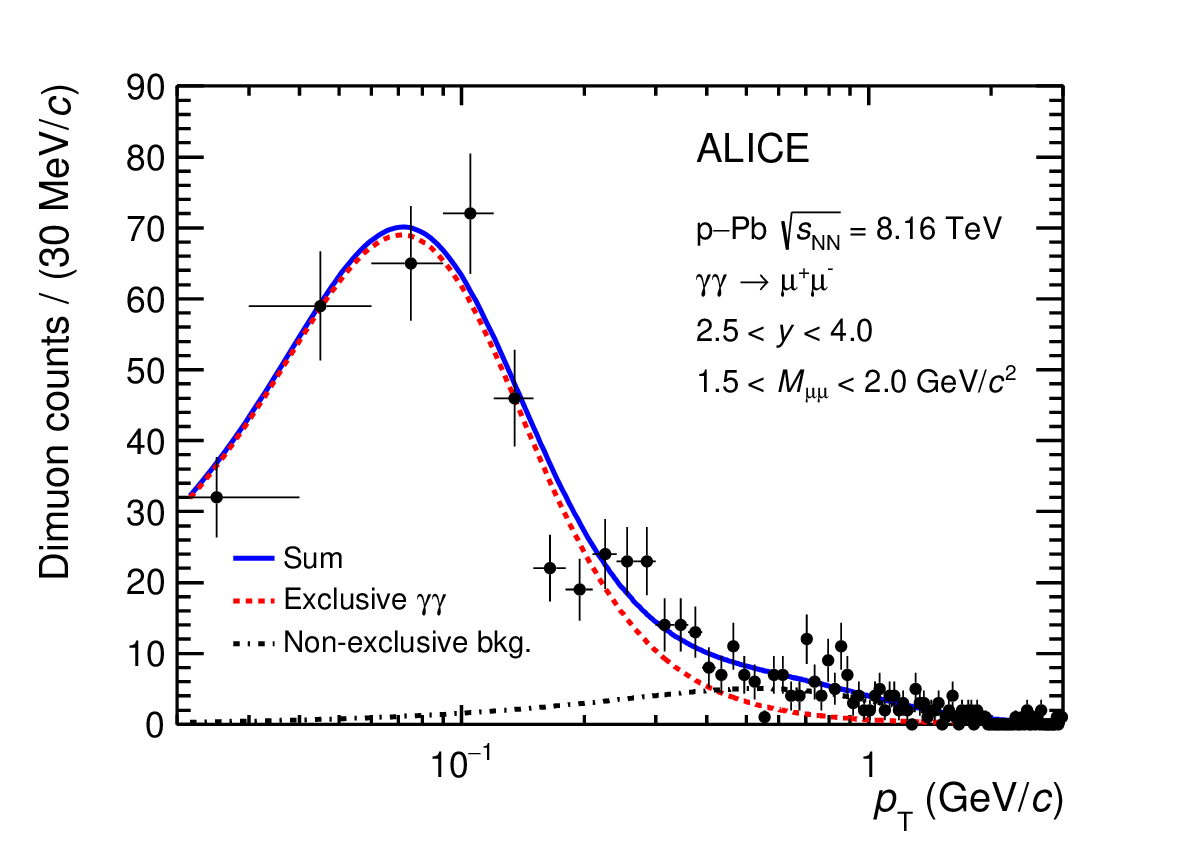}
 
    \end{center}
    \caption{Transverse momentum distribution of opposite-sign dimuons with $1.5 < \mmm  < 2.0$~\GeVmass and $2.5 < y < 4.0$. The data are represented by full circles with vertical bars for the statistical uncertainties and horizontal bars for the interval width.
    The solid line represents the fit to the data, and the dashed and dot-dashed lines represent the fit components.
 }
 \label{fig:landau}
\end{figure}

Figure~\ref{fig:landau} shows the \pt distribution of the dimuon candidates that satisfy the selections for $1.5 < \mmm < 2.0$~\GeVmass.
The data contain a mixture of exclusive and non-exclusive two-photon interactions, which are distinguished by their characteristic \pt distribution. While exclusive events dominate in the data at low \pt, the tail extending up to higher \pt is mostly due to non-exclusive interactions.

Exclusive \ggmm events are described with a Landau distribution, which is found to describe well the Monte Carlo data up to $\pt = 0.38$~\GeVc. A single-component fit of a Landau distribution is performed to the data requiring exclusivity on the proton side, as described in Sec.~\ref{sec:p-exc}, up to $\pt = 0.38$~\GeVc. In this selection, non-exclusive events are expected to be negligible. 
The location and scale parameters of the Landau distribution are extracted. The obtained location parameter ranges between 0.054 and 0.092 for the different bins. The scale parameter ranges between 0.23 and 0.39. They are then used when fitting the data that passed the standard selection, which includes both exclusive and non-exclusive components. 
Changing the maximum \pt value of the fitting interval or using a two-component model to account for exclusive and non-exclusive events instead of a single-component description might impact the location and scale parameters obtained from the fit of the exclusive-dominated sample. This is taken into account in the ``signal extraction'' systematic uncertainty (see Sec~\ref{sec:gg-syst}).

Non-exclusive events are modelled according to a parameterisation by H1 for dissociative events~\cite{H1:2013okq} with a function of the form $\mathrm{d}N/\mathrm{d}\pt \propto \pt \times \left( 1 + \pt^2 \times (b_{\rm dis}/n_{\rm dis})\right)^{-n_{\rm dis}}$, where $b_{\rm dis}$ and $n_{\rm dis}$ are free parameters. Non-exclusive dimuons represent $37\%$ of events in the analysed mass range and for $\pt<3$~\GeVc.

The $N_{\gamma \gamma}$ yields extracted from the fit are then  corrected for the acceptance and reconstruction efficiency $(\AxE)^{\gamma \gamma}$. The yields, the correction factors, and the cross sections are presented in Table~\ref{tab:crosssections-gg} for the different mass and rapidity intervals. The correction factors are evaluated by means of the Monte Carlo simulations introduced in Sec.~\ref{sec:MC-data}.

Additional activity in the \VZEROA, \ADA, \ZNA, or \SPD detectors results in event rejection and a corresponding correction needs to be applied.
Such events mainly originate from independent hadronic and electromagnetic pile-up processes.
The probability of event rejection due to pile-up of each veto is defined as the probability of detecting activity using events selected with an unbiased trigger based only on the timing of bunches crossing the interaction region. 
It is found to scale linearly with the expected number of collisions per bunch crossing. 
By varying the event selection in the analysis, the 
average pile-up probability varied from $3.7\%$ to $4.1\%$. Therefore, the pile-up probability is estimated as $p_{\rm pu} = (3.9 \pm 0.2)\%$ where most of the pile-up rejection ($3.7\%$) is from \VZEROA.
The average pile-up correction factor is calculated using $\epsilon_{\rm veto} = \exp(-p_{\rm pu} )$, and is found to be $\epsilon_{\rm veto} = (96.2 \pm 0.2)\%$.

\subsection{Signal extraction for \texorpdfstring{\jpsi}{jpsi} photoproduction candidates}

\begin{figure}[tb]
    \begin{center}
    \includegraphics[width=0.49\textwidth, trim={10 10 30 0}, clip]{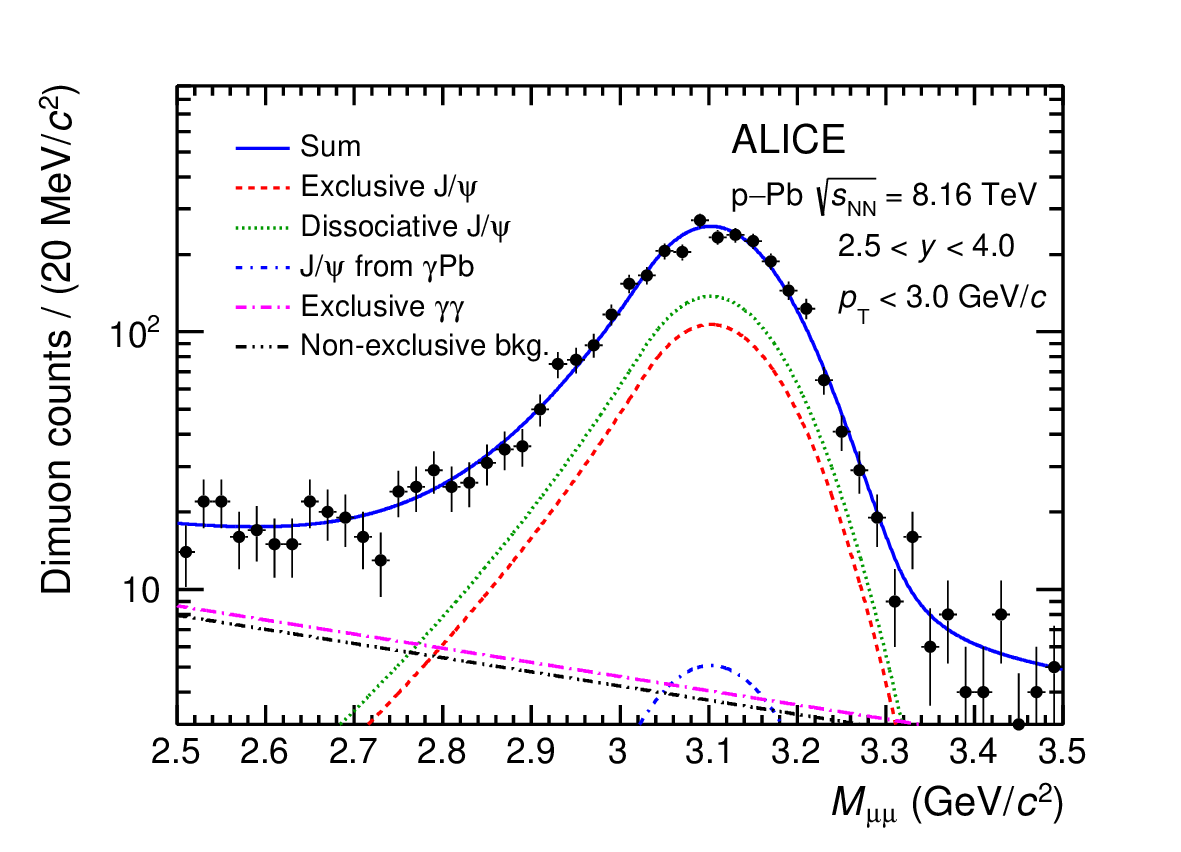}
    \hfill
    \includegraphics[width=0.49\textwidth, trim={10 10 30 0}, clip]{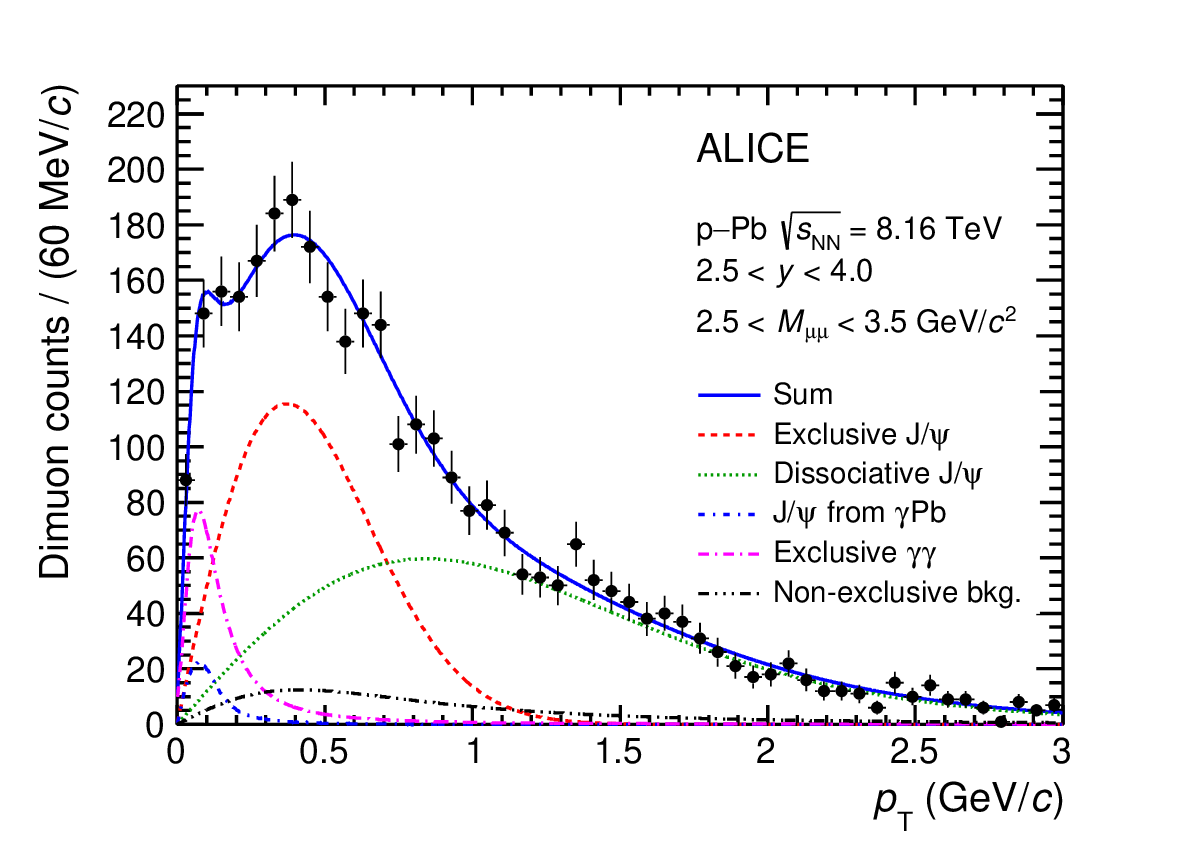}
    \end{center}
    \caption{Projections of the two-dimensional fit on the dimuon invariant mass (left) and \pt (right). }
    \label{fig:2d-jpsi}
\end{figure}

The yields of exclusive and dissociative \jpsi are obtained by performing an unbinned log-likelihood fit  to  dimuon invariant mass \mmm and transverse momentum \pt distributions simultaneously.
Events are selected in $2.5 < \mmm < 3.5$~\GeVmass and $\pt < 3$~\GeVc intervals. 
The dimuon invariant mass and \pt spectra after these selections are shown in Fig.~\ref{fig:2d-jpsi}. For the invariant mass distribution, the \jpsi peak 
is well described by a double-sided Crystal Ball parameterisation, which has a non-Gaussian tail at both sides of the resonance peak~\cite{Gaiser:1982yw, ALICE-PUBLIC-2015-006}. 
The \jpsi mass and its width at the pole position are free parameters of the fit, while the tail parameters in the Crystal Ball function are fixed to values obtained from fits to the 
Monte Carlo sample corresponding to the exclusive \jpsi photoproduction. The invariant mass distribution of the dimuon continuum is described by 
$\mathrm{d}N/\mathrm{d}\mmm \propto \exp(-a \mmm)$, where $a$ is a free parameter.

\jpsi events can be divided into three categories: exclusive photoproduction off protons, dissociative photoproduction off protons, and exclusive photoproduction off Pb nuclei. Dissociative photoproduction off Pb nuclei is vetoed by the \ZDC selection, as described in Sec.~\ref{sec:event}. 
The events contained in the dimuon continuum below the \jpsi peak can either be exclusive or non-exclusive two-photon interactions.
The various physics processes in the \jpsi peak and in the dimuon continuum can be distinguished by their different \pt distributions. 

\begin{figure}[tb]
    \begin{center}
    \includegraphics[width=0.49\textwidth, trim={10 10 30 0}, clip]{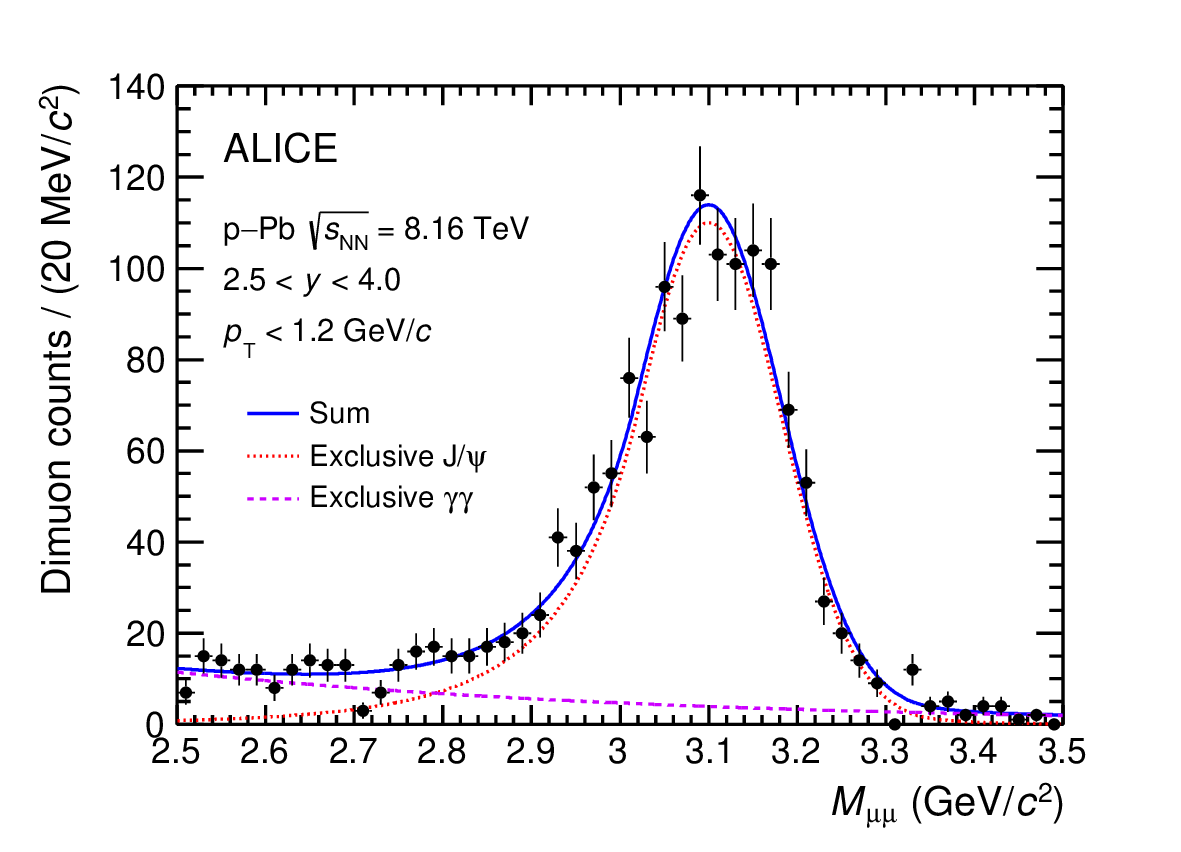}
    \includegraphics[width=0.49\textwidth, trim={10 10 30 0}, clip]{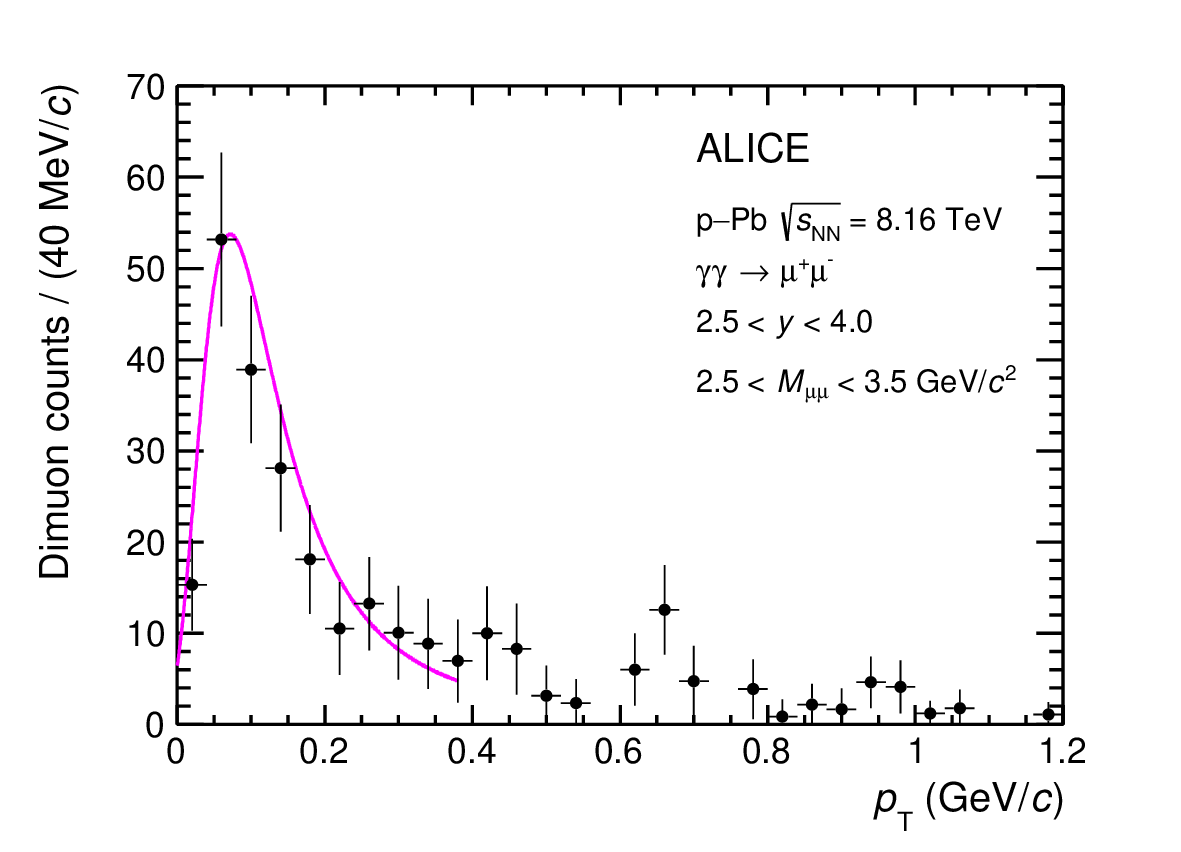}
    \end{center}
    \caption{
    Left: dimuon invariant mass distribution using the selection given in Sec.~\ref{sec:p-exc}, fitted with a two-component model to separate \jpsi events from two-photon interactions in the continuum. 
    Right: \pt distribution of the exclusive $\ggmm$ continuum extracted using the sPlot technique. The distribution is  fitted with a Landau distribution. 
\label{fig:splot}}
\end{figure}

The \pt distribution for \ggmm events below the \jpsi peak is modelled with a Landau distribution for which the location and scale parameters are fixed, similarly as in Sec.~\ref{sec:gg-sig}.
Their values are obtained using the sample described in Sec.~\ref{sec:p-exc}.  
In order to factorise the \pt distribution of \jpsi and continuum dimuon events, the numerical tool sPlot is used~\cite{Pivk:2004ty}. 
Based on an extended maximum likelihood fit to the mass distribution of the sample (left panel of Fig.~\ref{fig:splot}), the sPlot procedure assigns weights denoted as $sw_n$ on an event-by-event basis. 
Assuming these weights can be computed as a linear combination of conditional probabilities, they are given by the following formula for the category $n =1, 2$ of events in the sample (\jpsi signal or \ggmm)
\begin{equation}
    sw_n(\mmm) = \frac{\sum_{i=1}^{N_s} V_{ni}f_i(\mmm)}{\sum_{j=1}^{N_s} N_{j}f_j(\mmm)}
    \text{,}
\end{equation}
where $f$ is the probability density function of the fit, \mmm denotes the mass used as the discriminating variable for each event, $i$ and $j$ are the indices indicating a sum over the $N_s = 2$ categories, and $V$ is the covariance matrix of the yields $N_j$ which is evaluated in a separate fit, in which all shape-related parameters are fixed. 
The \pt distribution for two-photon interactions extracted with the sPlot technique is shown in the right panel of Fig.~\ref{fig:splot} and is fitted up to $\pt = 0.38$~\GeVc with a single-component fit parametrised with a Landau distribution, from which the location and scale parameters are extracted.
The small correlation between the mass and \pt of dimuons produced in two-photon interactions was found to have a negligible impact on the sPlot procedure. 
In addition, the extracted number of \ggmm events in the \jpsi peak range ($2.5 < \mmm < 3.5$~\GeVmass) 
is compared with \starlight and an agreement within $1 \sigma$ is found (accounting for the statistical uncertainties only). This number is also in good agreement with the number of continuum dimuon events extracted from the final two-dimensional fit.

\begin{figure}
  \includegraphics[width=0.49\textwidth, trim={10 10 30 0}, clip]{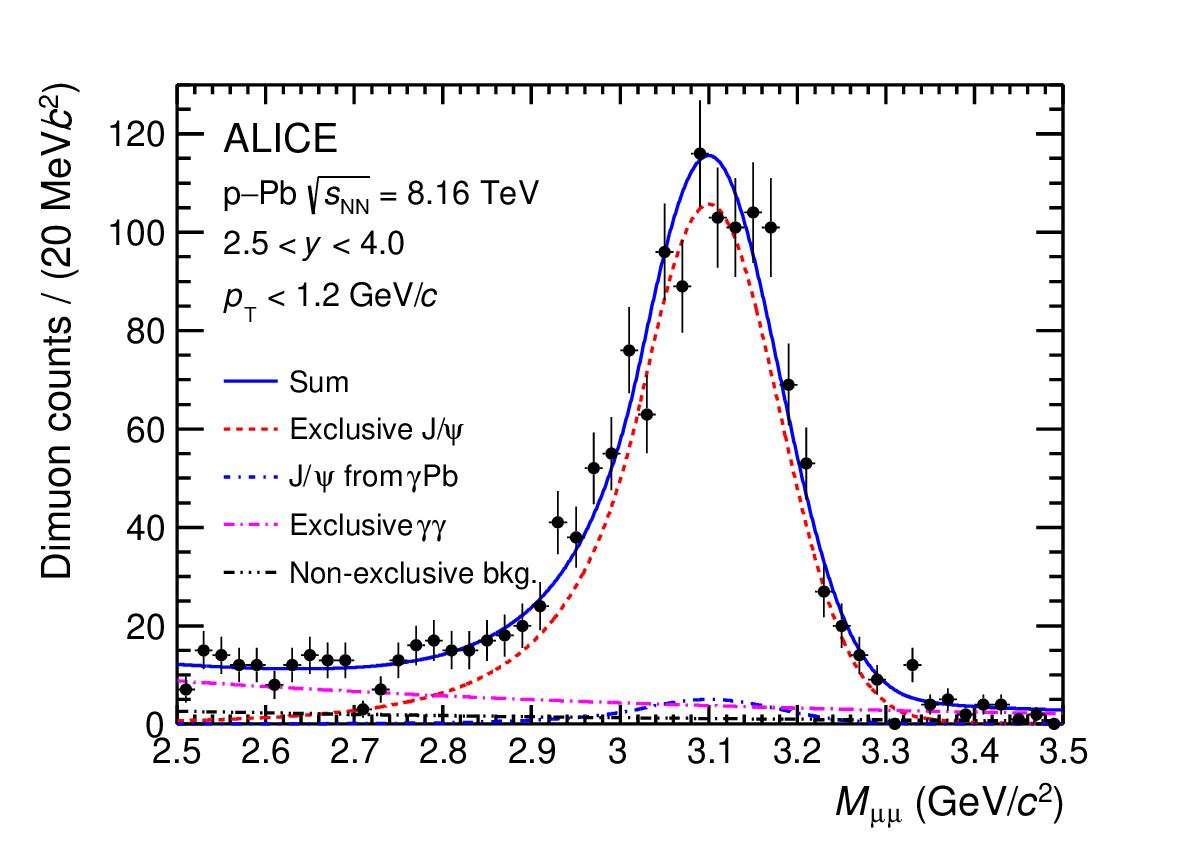}
  \includegraphics[width=0.49\textwidth, trim={10 10 30 0}, clip]{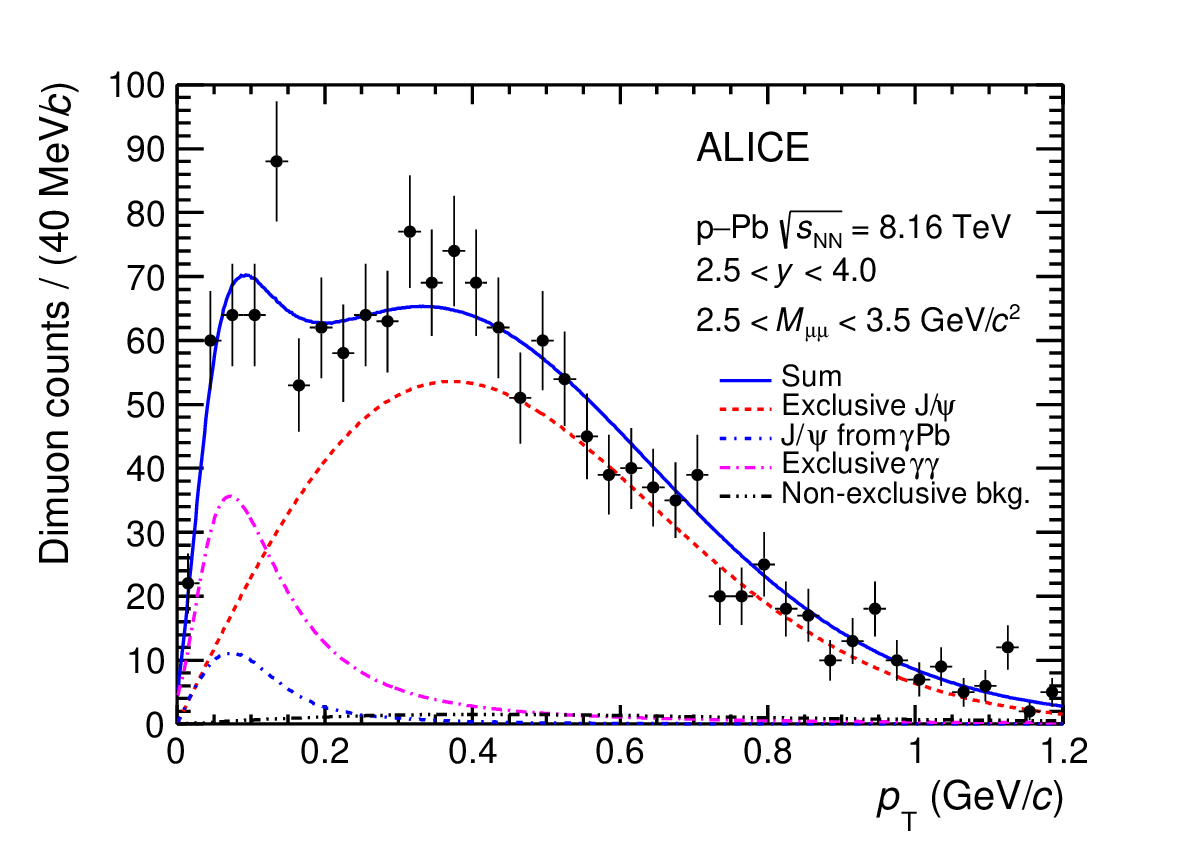}
  \caption{Projections of the two-dimensional fit on the dimuon invariant mass (left) and \pt (right) with the selection described in Sec.~\ref{sec:p-exc} to extract the shape of the \pt distribution for the exclusive $\jpsi$ events in \gp interactions. }
  \label{fig:exclusiveonly}
  \end{figure}

The shape of the \pt distribution for the exclusive $\jpsi$ events in \gp interactions is given by the H1 parameterisation~\cite{H1:2013okq} $\mathrm{d}N/\mathrm{d}\pt \propto \pt \times \exp(-b_{\rm exc} \pt^2 )$, where $b_{\rm exc}$ is a fixed parameter. \jpsi mesons coming from \psip decays are also included in this contribution.
The $b_{\rm exc}$ value is determined using the sample described in Sec.~\ref{sec:p-exc}, by fitting simultaneously the dimuon invariant mass and \pt without the contribution of dissociative \jpsi events. The dimuon invariant mass and \pt projections of this fit are shown in Fig.~\ref{fig:exclusiveonly}. 
Studies conducted with the RAPGAP Monte Carlo program~\cite{Jung:1993gf} in the kinematic range of the present measurement show that more than 99\% of dissociative \jpsi events are removed by the selection requiring exclusivity on the proton side.
The values obtained are 
$b_{\rm exc} = 3.62 \pm 0.14$~$[\text{GeV/c}\,\,]^{-2}$ for $2.5 < y_{\mu\mu} < 4.0$, 
$b_{\rm exc} = 3.38 \pm 0.17$~[\GeVc]$^{-2}$ for $3.25 < y_{\mu\mu} < 4.0$ ($27 < \Wgp < 39$~\GeV), and 
$b_{\rm exc} = 3.86 \pm 0.20$~[\GeVc]$^{-2}$ for $2.5 < y_{\mu\mu} < 3.25$ ($39 < \Wgp < 57$~\GeV).
The \pt resolution of the muon spectrometer is the main limitation in unfolding these values and comparing them with the H1 measurement of the $t$-slope, 
$b_{\rm exc} = (4.3 \pm 0.2)$~[\GeV]$^{-2}$ for $25 < \Wgp < 80$~\GeV~\cite{H1:2013okq}.
As an alternative method to extract the $b_{\rm exc}$ value, the \jpsi \pt distribution obtained with sPlot was fitted with a two-component model including \jpsi events from \gp interactions and from \gPb interactions. The bias induced by the method used to extract $b_{\rm exc}$ is accounted for in the ``signal extraction'' systematic uncertainty (see Sec.~\ref{sec:syst-jpsi}).

The \pt distribution for coherent \jpsi photoproduction in \gPb interactions is obtained using the corresponding reconstructed Monte Carlo sample within the specified mass and \pt ranges.
The \pt distributions for dissociative \jpsi events and non-exclusive two-photon interactions are modelled by functions of the form $\mathrm{d}N/\mathrm{d}\pt \propto \pt \times \left( 1 + \pt^2 \times (b_{\rm dis}/n_{\rm dis})\right)^{-n_{\rm dis}}$ where $b_{\rm dis}$ and $n_{\rm dis}$ are free parameters. 

Five parameterisations for exclusive \jpsi photoproduction off protons, dissociative \jpsi photoproduction off protons, \jpsi photoproduction off Pb nuclei, exclusive, and non-exclusive \ggmm, are defined as products of each corresponding mass and \pt distributions.
The normalisation for the component corresponding to \jpsi produced in \gPb interactions is fixed to the expected number according to a computation based on the measurement from Ref.~\cite{Acharya:2019vlb} under the assumption that the fraction of low- and high-energy photon contributions to the forward rapidity measurement is the same as predicted by \starlight.
The normalisation for all other components are free parameters of the fit.

The extracted yields of exclusive and dissociative \jpsi from \gp interactions 
are corrected for acceptance and reconstruction efficiency $(\AxE)^{\jpsi}$, which are obtained from the Monte Carlo simulation samples described in Sec.~\ref{sec:MC-data}, having values ranging  from 18\% to 21\%. 

The extracted yields are corrected for the feed-down contribution of \jpsi mesons coming from \psip decays, denoted \fD. 
Following the procedure described in Ref.~\cite{ALICE:2013wjo}, 
\fD is given by
\begin{equation}
\displaystyle
    \fD = \frac{ \sigma \left(\psip\right) \times BR(\psip \rightarrow \jpsi + X) \times (\AxE)_{\jpsi}^{\rm FD}}{\sigma (\jpsi)  \times (\AxE)_{\jpsi}}
    \text{,}
\end{equation}
where $\sigma(\jpsi)$ and $\sigma \left(\psip\right)$ are the cross sections of \jpsi and \psip productions, respectively, at a given rapidity, the branching ratio for the decay of a $\psip$ to \jpsi is $BR(\psip \rightarrow \jpsi + X) = (61.4 \pm 0.6) \%$~\cite{ParticleDataGroup:2020ssz}, and $(\AxE)_{\jpsi}$ and $(\AxE)_{\jpsi}^{\rm FD}$ are the acceptance and reconstruction efficiency for events with a \jpsi produced directly from \gp interactions and from \psip decays, respectively. In order to compute \fD, the ratio $ \sigma \left(\psip\right)/\sigma (\jpsi) = 0.150 \pm 0.013 \text{ (stat.)}\pm 0.011 \text{ (syst.)}$ is taken from the H1 measurement for $40 < \Wgp < 70$~\GeV~\cite{H1:2002yab}. 
The $(\AxE)_{\jpsi}^{\rm FD}$ values  are evaluated under the assumption that feed-down \jpsi mesons inherit the transverse polarisation of their $\psip$ parents, as indicated by previous measurements~\cite{BES:1999guu}.
The obtained \fD values range between $(9.1 \pm 1.2)\%$ and $(9.3 \pm 1.2)\%$ depending on the rapidity interval. The uncertainties are obtained by summing the statistical and systematic uncertainties of the H1 measurement and branching ratio uncertainties in quadrature.
Finally, the numbers are corrected for pile-up, as discussed in Sec.~\ref{sec:gg-sig}.

\subsection{Systematic uncertainties}

The experimental systematic uncertainties for the exclusive dimuon  production from two-photon interactions and for the photoproduced \jpsi are listed in Table~\ref{tab:syst}. 
The systematic sources can be divided into three types: those common to both measurements and   those affecting one or the other. 

\subsubsection{Systematic uncertainties common to both measurements}

The uncertainty on the integrated luminosity is discussed in Sec.~\ref{sec:trigger}, and amounts to 1.8\%.
The systematic uncertainties on muon trigger efficiency, tracking efficiency, and muon matching efficiency were obtained as described in Ref.~\cite{ALICE:2018mml}. 
The single-muon trigger response functions evaluated in data and Monte Carlo simulations are incorporated in the acceptance and efficiency $(\AxE)$ calculations for the reconstruction of the dimuons. The differences between $(\AxE)$ calculations when incorporating the response functions either from data or Monte Carlo range from 0.1\% to 4.9\% depending on the studied process and rapidity interval. The total uncertainty is obtained by combining this contribution in quadrature with the uncertainty on the intrinsic efficiency of muon trigger detectors, which amounts to 1\%.

The uncertainty on the tracking efficiency was calculated by comparing the efficiencies evaluated in data and Monte Carlo simulations. 
These efficiencies are calculated according to the tracking algorithm by combining the efficiency of each tracking plane measured using the redundancy of the system.
The estimated value of the systematic uncertainty related to the tracking efficiency is 1\% in this data sample.
The muon matching efficiency is the efficiency of associating a muon track candidate to a trigger track above the $0.5$~\GeVc \pt threshold in the trigger chambers of the muon spectrometer.
Its uncertainty is estimated by varying the $\chi^2$ cutoff applied to the pairing of the reconstructed tracks in the muon tracking and triggering systems, and it is found to be 1\%.

The pile-up correction factor, discussed in Sec.~\ref{sec:gg-sig}, has a relative uncertainty of 0.2\%.

The uncertainty on the veto efficiency of the \VZEROC is calculated by varying the number of allowed cells with a signal over the threshold in the offline selection.  When increasing this number, the numbers of exclusive \jpsi and
\ggmm events are found to be stable, while the number of dissociative \jpsi events increases, as
the sample is more sensitive to contamination from inclusive photoproduction or hadronic production of \jpsi mesons
which have a similar behaviour in \pt. The expected number of dissociative \jpsi events is computed as
the number of exclusive \jpsi events multiplied by the ratio of dissociative-to-exclusive \jpsi events
when all the fired cells in \VZEROC are required to be matched to a muon. The systematic uncertainty on
the number of dissociative \jpsi events is computed as the relative difference between the expected and
extracted numbers of dissociative \jpsi events and is found to be 12.7\%, while the systematic uncertainty
on the number of exclusive \jpsi events is obtained by varying the condition on \VZEROC and the obtained
value is 2.6\%. Similarly, the uncertainty on the number of exclusive \ggmm events is obtained by
varying the condition on \VZEROC (see line “V0C veto” in Table~\ref{tab:syst}) and the obtained values vary between
0.5\% and 1.7\%.

\subsubsection{Uncertainties associated with the dimuon continuum production}
\label{sec:gg-syst}

The main source of systematic uncertainty on the \ggmm signal extraction is obtained by varying both parameters of the Landau distribution within their statistical uncertainties obtained 
from fitting the purely exclusive sample described in Sec.~\ref{sec:p-exc} and taking into account their correlation (see line ``signal extraction'' in Table~\ref{tab:syst}). 

In the lowest invariant mass interval studied, $1.0 < \mmm < 1.5$~\GeVmass, the production of $\phi$ mesons decaying to dimuons might contaminate the sample. 
The expected number of $\phi \rightarrow \mu^+ \mu^-$ events in the sample at low mass, $N_\phi$, is computed. The calculation is based on the cross section ratio of $\phi$ photoproduction with respect to \jpsi~production based on \starlight and their branching ratios provided by the PDG~\cite{ParticleDataGroup:2020ssz}, detector acceptance and efficiency factors, and the number of \jpsi mesons measured in the muon spectrometer.
The uncertainty induced by this contamination is estimated by comparing $N_\phi$ to the number of \ggmm events. It is found to be $1.5 \%$.

\subsubsection{Uncertainties associated with  the \texorpdfstring{\jpsi}{jpsi} photoproduction only}
\label{sec:syst-jpsi}

The main source of systematic uncertainty on the \jpsi signal extraction is obtained by varying the $b_{\rm exc}$ parameter within its statistical uncertainty determined from fitting the purely exclusive sample described in Sec.~\ref{sec:p-exc} (see line ``signal extraction'' in Table~\ref{tab:syst}).
It ranges between 2.9\% and 5.5\%. Changing the \pt model for the exclusive \ggmm component and varying the number of \jpsi events produced in \gPb interactions were found to have
a negligible impact on signal extraction.

The photon flux, which enters in the computation of the cross section presented in Sec.~\ref{sec:x-sec-jpsi}, is computed using \starlight. Its uncertainty is obtained by varying the nuclear radii and the nuclear density $\rho_0$ of the Pb nucleus, assuming that the latter has a cubic dependence on the radius. The radius of the lead nucleus is changed by $\pm 0.5$ fm, which corresponds to the nuclear skin thickness. This uncertainty is evaluated to be $2 \%$. The branching ratio of \jpsi decaying into dimuons and its uncertainty (0.55\%) are given by the Particle Data Group~\cite{ParticleDataGroup:2020ssz}.

For the measured ratio of dissociative-to-exclusive cross sections, $\sigma^{\rm diss}/\sigma^{\rm exc}$, most of the systematic uncertainties cancel out. The remaining sources of uncertainty are due to the variation of the $b_{\rm exc}$ parameter, and the variation on the number of allowed fired \VZEROC cells. 
The systematic uncertainties on the ratio given in Table~\ref{tab:syst} are then computed as the quadratic sum of these two components only.

\section{Results}

\subsection{Cross sections for the dimuon continuum in two-photon interactions}

The cross section corresponding to the exclusive \ggmm process is measured using
\begin{equation}
\displaystyle
\frac{\mathrm{d} \sigma^{\gamma\gamma}}{\mathrm{d}\mmm}({\rm p+Pb} \rightarrow {\rm p+Pb}+\mu^+ + \mu^-)  = \frac{N_{\gamma \gamma}}{(\AxE)^{\gamma\gamma} \times \mathcal{L} \times \epsilon_{\rm veto} \times \Delta \mmm} \text{,}
\end{equation}
where $N_{\gamma \gamma}$ is the number of reconstructed \ggmm events, $(\AxE)^{\gamma\gamma}$ is the corresponding factor which takes into account acceptance and reconstruction efficiency in the mass and rapidity interval studied, $\epsilon_{\rm veto}$ 
is the pile-up correction factor and 
$\Delta \mmm$ is the width of the invariant mass interval.

The rapidity range of the experimental results corresponds to a high-energy photon emitted from the proton (corresponding to small impact parameters with respect to the proton) and a low-energy photon emitted from the nucleus  (corresponding to large impact parameters with respect to the nucleus). The differential cross sections, $\mathrm{d}\sigma^{\gamma\gamma}/\mathrm{d}\mmm $, are presented in Table~\ref{tab:crosssections-gg} in two rapidity intervals and integrated over rapidity along with 
the predictions from \starlight 2.2.0 and \superchic 4.15~\cite{Harland-Lang:2020veo} for comparison. 

The \starlight generator simulates UPCs at colliders based on the equivalent photon approximation.  \superchic was designed for exclusive production in proton--proton collisions and has been extended to collisions involving nuclei starting from Ref.~\cite{Harland-Lang:2018iur}.  For $\gamma\gamma$-induced dilepton production, \superchic provides calculations on amplitude level to treat the  probability of no hadronic interaction within the same collision. Both generators implement LO QED calculations, neglecting final-state radiation. 

\begin{figure}[tb]
\begin{center}
\includegraphics[width=0.49\textwidth]{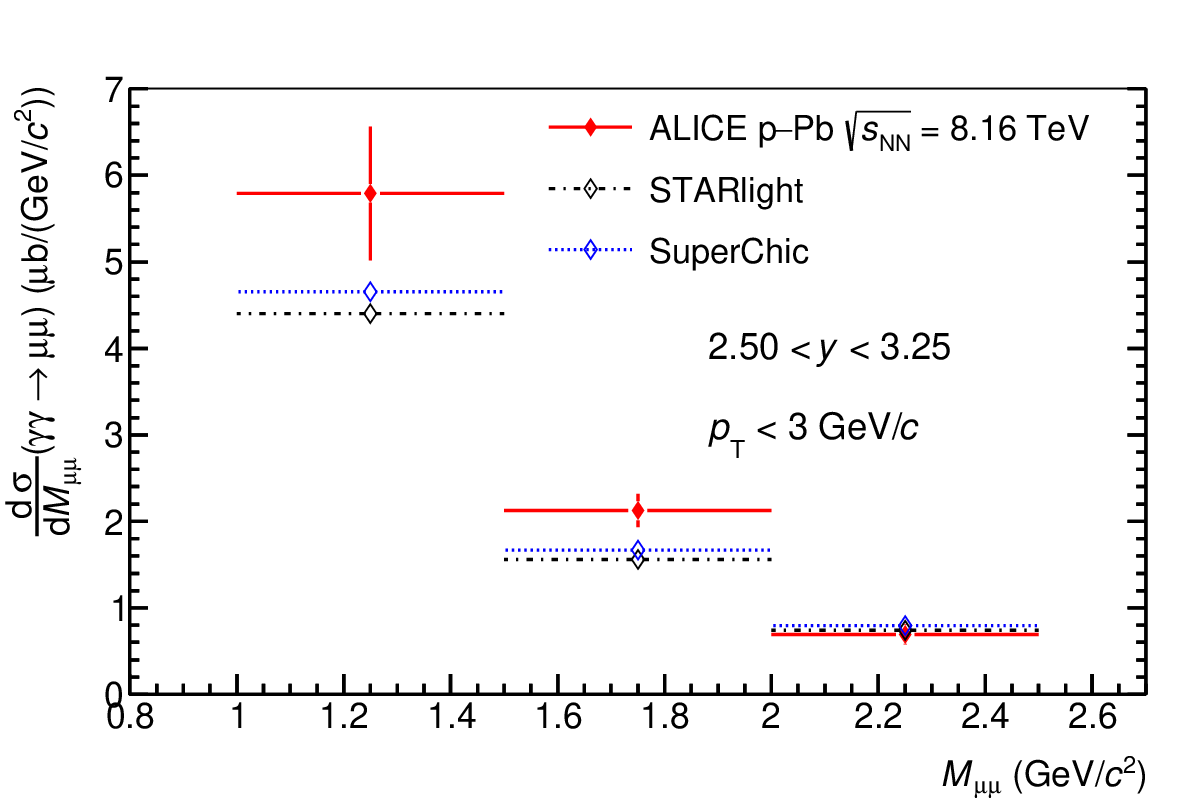}
\includegraphics[width=0.49\textwidth]{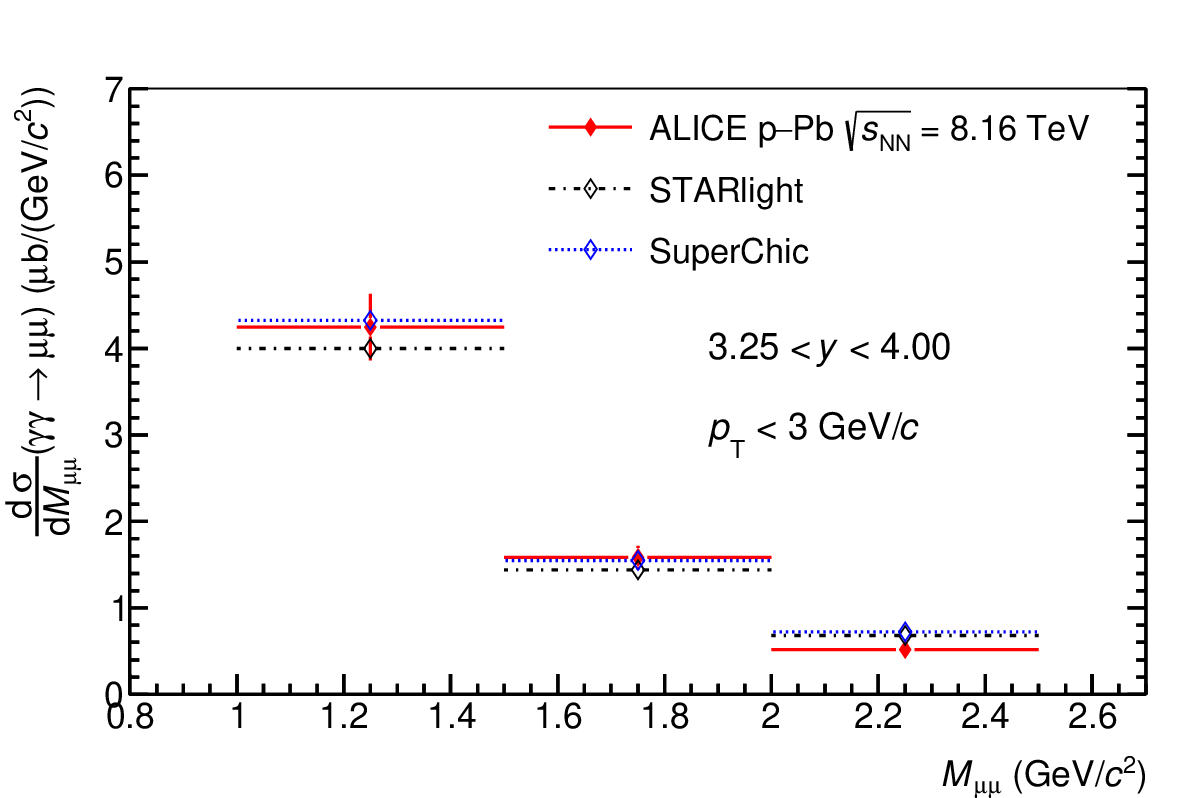}
\end{center}
\caption{Differential cross sections for exclusive $\ggmm$ production measured by ALICE in \pPb UPCs at \eightsixteen, as a function of \mmm, for $2.5 < y < 3.25$ (left) and $3.25 < y < 4$ (right). The vertical error bars represent the statistical and systematic uncertainties summed in quadrature. The results are compared with the prediction from \starlight~\cite{Klein:1999qj, Klein:2016yzr} and from \superchic~\cite{Harland-Lang:2020veo}.}
\label{fig:gg-crossX}
\end{figure}

 The measured cross sections and predictions from \starlight and \superchic are shown in Fig.~\ref{fig:gg-crossX}. 
Both predictions agree within $3$ standard deviations, depending on the mass and rapidity intervals. In the two lowest mass intervals, the central values of the measured cross sections are larger compared with \starlight and \superchic, while the opposite behaviour is seen in the highest mass interval. For the kinematic intervals studied, \superchic predicts larger cross sections than \starlight. 
 The difference between \starlight and \superchic discussed in Ref.~\cite{Harland-Lang:2021ysd}, related to the sharp cut-off on the impact parameter between the produced dilepton and the nucleus, is  found  not to be the primary source of discrepancy observed here. 

The relative uncertainties on the measurements vary from 7\% to 17\%. This is significantly larger than the 2\% uncertainty for the photon flux used in the calculation of the photoproduction cross section presented in Sec.~\ref{sec:x-sec-jpsi}. Thus, with the current experimental precision, it is not possible to constrain  the photon fluxes via the \ggmm measurement. 

\begin{table}[tb]
\centering
\caption{Differential cross sections $\mathrm{d}\sigma^{\gamma \gamma}/\mathrm{d}\mmm$ for exclusive $\ggmm$ production in \pPb UPCs at \eightsixteen for each mass and rapidity interval, measured by ALICE and computed with \starlight and \superchic. The first uncertainty is the statistical one and the second uncertainty is the systematic one. The corresponding number of exclusive \ggmm events with their statistical uncertainties and factors of acceptance times reconstruction efficiency are given.}
\label{tab:crosssections-gg}
\small
\begin{tabular}{ccccccc}
\hline
    \multirowcell{3}{Mass range \\ (\GeVmass)} & \multirowcell{3}{Rapidity range}    & \multirowcell{3}{$N_{\gamma \gamma}$} & \multirowcell{3}{$(\AxE)$} &   \multirowcell{3}{$\mathrm{d}\sigma^{\gamma \gamma}/\mathrm{d}\mmm$ \\ ({\textmu}b $c^2/$GeV)} & \multirowcell{2}{$\mathrm{d}\sigma^{\gamma \gamma}/\mathrm{d}\mmm$ \\ ({\textmu}b $c^2/$GeV) \\ (\starlight)} & \multirowcell{3}{$\mathrm{d}\sigma^{\gamma \gamma}/\mathrm{d}\mmm$ \\ ({\textmu}b $c^2/$GeV)  \\ (\superchic)}\\
    & & & & & \\
        & & & & & \\

    \hline
   \multirowcell{3}{(1.0, 1.5)} & \multirowcell{1}{(2.5, 4)} & $618 \pm 33$ & $1.66\%$ & $9.84 \pm 0.52 \pm 0.49$ & $8.45$ & $8.98$  \\
      & \multirowcell{1}{(3.25, 4)} & $522 \pm 31$ & $3.23\%$ & $4.26 \pm 0.25 \pm 0.20$ & $4.05$ & $4.33$ \\
     & \multirowcell{1}{(2.5, 3.25)} & $99 \pm 11$ & $0.45\%$ & $5.75 \pm 0.67 \pm 0.34$ & $4.39$ & $4.65$ \\
    \hline
     \multirowcell{3}{(1.5, 2.0)} & \multirowcell{1}{(2.5, 4)} & $437 \pm 26$ & $3.04\%$ & $3.79 \pm 0.22 \pm 0.20$  & $3.00$ & $3.22$ \\
     & \multirowcell{1}{(3.25, 4)} & $283 \pm 19$ & $4.74\%$ & $1.58 \pm 0.12 \pm 0.09$ & $1.44$ & $1.55$ \\
    & \multirowcell{1}{(2.5, 3.25)} & $150 \pm 14$ & $1.82\%$ & $2.17 \pm 0.20 \pm 0.15$ & $1.56$ & $1.67$ \\
    \hline
    \multirowcell{3}{ (2.0, 2.5)} & \multirowcell{1}{(2.5, 4)} & $191 \pm 18$ & $4.09\%$ & $1.23 \pm 0.12 \pm 0.07$ & $1.42$ & $1.52$ \\
    & \multirowcell{1}{(3.25, 4)} & $103 \pm 13$ & $5.32\%$ & $0.511 \pm 0.065 \pm 0.034$ & $0.673$ & 0.724 \\
     & \multirowcell{1}{(2.5, 3.25)} & $85 \pm 13$ & $3.25\%$ & $0.692 \pm 0.101 \pm 0.060$ & $0.744$ & 0.794 \\
    \hline
\end{tabular}
\end{table}

\begin{table}[tb]
\centering
\caption{Summary of systematic uncertainties on the measured cross sections. The value ranges correspond to different rapidity intervals.
Uncertainties on signal extraction, tracking, trigger, and muon matching efficiencies are considered as uncorrelated across $y$. All other components are taken as fully correlated across the rapidity $y$. The final uncertainties for $\gamma \gamma$ and \jpsi, labeled ``Total'', are obtained as the sum in quadrature of common uncertainties and those affecting one signal or the other. 
}
\label{tab:syst}
\begin{tabular}{|l|lll|}
\hline
\multirowcell{1}{Signal} & \multirowcell{1}{Source}     & \multirowcell{1}{Mass range (\GeVmass)} &  \multirowcell{1}{ Value } \\
\hline
\multirowcell{6}{All} & Luminosity               & &      1.8\%     \\
& Tracking efficiency       & &      1\%     \\
& Matching efficiency                & &      1\%     \\
& Pile-up correction  & &      0.2\%   \\
\cline{2-4}
& \textbf{Total common}                   & &    \textbf{2.3\%}  \\ \hline 
\multirowcell{13}{$\gamma \gamma$ only} &  & \multirowcell{1}{ (1.0, 1.5)} & from 2.1\% to 3.4\% \\ 
& Muon trigger efficiency & \multirowcell{1}{ (1.5, 2.0)} & from 2.5\% to 5.0\% \\ 
& & \multirowcell{1}{ (2.0, 2.5)} & from 1.6\% to 3.3\% \\ 
\cline{2-4}
 & $\phi \rightarrow \mu^+ \mu^-$ contamination & \multirowcell{1}{(1.0, 1.5)} & 1.5\%  \\
\cline{2-4}
&  & \multirowcell{1}{ (1.0, 1.5)} & 1.2\% \\ 
& \VZEROC veto & \multirowcell{1}{ (1.5, 2.0)} & 1.7\% \\ 
& & \multirowcell{1}{ (2.0, 2.5)} & 0.5\% \\ 
\cline{2-4}
 & & \multirowcell{1}{(1.0, 1.5)} & from $3.2\%$ to $3.9\%$ \\
   &  Signal extraction &   \multirowcell{1}{ (1.5, 2.0)} & from $3.3\%$ to $4.4\%$ \\
&  &   \multirowcell{1}{(2.0, 2.5)} & from $4.9\%$ to $7.6\%$   \\
   \cline{2-4}
& & \multirowcell{1}{(1.0, 1.5)} & \textbf{from 4.9\% to 6.0\%}    \\
& \textbf{Total} & \multirowcell{1}{(1.5, 2.0)} & \textbf{from 5.5\% to 7.1\%} \\
 & & \multirowcell{1}{(2.0, 2.5)} & \textbf{from 6.0\% to 8.6\%} \\
 \hline
\multirowcell{9}{\jpsi only} &  Muon trigger efficiency &  & $1.1\%$ \\
 & Branching ratio          & &     0.55\%     \\
 & Photon flux              & &     2\%     \\
 & $\delta(1+\fD)$          & &   1.1\% \\

& \VZEROC veto & & 2.6\% (excl.), 12.7\% (diss.)   \\
& \multirowcell{2}[0pt][l]{Signal extraction} & \multirowcell{2}{(2.5, 3.5)} &   from 3.6\% to 5.5\% (excl.), \\
  & &  &   from 2.9\% to 4.4\% (diss.) \\
 \cline{2-4}
 & \multirowcell{2}[0pt][l]{\textbf{Total} }       &     &    \textbf{from 5.6\%  to 7.0\% (excl.),} \\
  &  &     &   \textbf{from 13.5\% to 13.9\% (diss.)} \\
\hline
\multirowcell{3}{$\displaystyle \frac{\sigma^{\rm diss}}{\sigma^{\rm exc}}$} & \VZEROC veto & & 12.7\% \\
 & Signal extraction & & from 6.2\% to 7.6\% \\
 \cline{2-4}
 & \textbf{Total} & & \textbf{from 14.1\% to 14.8\%} \\
 \hline
\end{tabular}
\end{table}

\subsection{Cross sections for \texorpdfstring{\jpsi}{jpsi} photoproduction off protons}
\label{sec:x-sec-jpsi}

The cross sections corresponding to exclusive and dissociative \jpsi photoproduction off protons are measured using
\begin{equation}
\displaystyle
\frac{\mathrm{d} \sigma}{\mathrm{d}y} ({\rm p+Pb} \rightarrow {\rm p^{(*)}+Pb}+\jpsi) = \frac{N_{\jpsi}}{(\AxE)^{\jpsi} \times (1+\fD) \times \mathcal{L} \times \epsilon_{\rm veto} \times BR \times \Delta y}
\text{,}
\end{equation}
where $N_{\jpsi}$ is the number of reconstructed exclusive or dissociative \jpsi in the dimuon decay channel, $(\AxE)^{\jpsi}$ is the corresponding factor of acceptance times reconstruction efficiency in the rapidity interval studied, and $BR = (5.961 \pm 0.033)\%$ is the branching ratio for the decay into a muon pair~\cite{ParticleDataGroup:2020ssz}.

The cross section ${\rm d} \sigma/{\rm d}y ({\rm p+Pb} \rightarrow {\rm p^{(*)}+Pb}+\jpsi)$ is related to the \gp cross section, $\sigma(\gamma+{\rm p} \rightarrow \jpsi+{\rm p^{(*)}})$, through the photon flux $\mathrm{d} n/\mathrm{d} k$,
\begin{equation}
    \displaystyle
    \frac{{\rm d} \sigma}{{\rm d}y}({\rm p+Pb} \rightarrow {\rm p^{(*)}+Pb}+\jpsi) = k \frac{\mathrm{d}n}{\mathrm{d}k}
\sigma(\gamma+{\rm p} \rightarrow \jpsi+{\rm p^{(*)}})
\text{.}
\end{equation}

Here, $k$ is the photon energy, which is determined by the \jpsi mass and rapidity, $k = (1/2)M_{\jpsi} \exp{ (-y)}$.
The photon flux is calculated using \starlight in impact parameter space and convoluted with the probability of no hadronic interaction. The average photon flux values for the different rapidity intervals are listed in Table~\ref{tab:crosssections-jpsi}, together with the extracted cross sections $\sigma(\gamma+ {\rm p} \to \jpsi+ {\rm p})$ and $\sigma(\gamma+ {\rm p} \to \jpsi+ {\rm p}^{(*)})$ and the corresponding $\langle \Wgp \rangle$. The latter is computed as the average of \Wgp weighted by the cross section $\sigma(\gp)$ from \starlight.

\begin{table}[tb]
\centering
\caption{Rapidity differential cross sections $\mathrm{d}\sigma^{\rm exc}_{\jpsi}/\mathrm{d}y$ and $\mathrm{d}\sigma^{\rm diss}_{\jpsi}/\mathrm{d}y$ and the corresponding cross sections $\sigma(\gamma+{\rm p} \rightarrow \jpsi+ {\rm p})$ and $\sigma(\gamma+{\rm p} \rightarrow \jpsi+ {\rm p}^{(*)})$ for exclusive and dissociative \jpsi photoproduction off protons in \pPb UPCs at \eightsixteen for each rapidity range. The first uncertainty is the statistical one and the second uncertainty is the systematic one. The numbers of events obtained from signal extraction with their statistical uncertainties, $N_{\jpsi}^{\rm exc}$ and $N_{\jpsi}^{\rm diss}$, the photon flux, and the range and the mean of $\Wgp$ are also presented.}
\label{tab:crosssections-jpsi}
\small
\begin{tabular}{lllllll}
\hline
\\[-8pt]
    \multirowcell{2}{ Rapidity range }   & \multirowcell{1}{$N_{\jpsi}^{\rm exc}$,} & \multirowcell{1}{$\mathrm{d}\sigma^{\rm exc}_{\jpsi}/\mathrm{d}y$,} &  \multirowcell{2}{$k \mathrm{d}n/\mathrm{d}k$}   & \multirowcell{2}{$\Wgp$ \\ (GeV)}    &  \multirowcell{2}{$\langle \Wgp \rangle$ \\ (GeV)} & \multirowcell{1}{$\sigma(\gamma+{\rm p} \rightarrow \jpsi+{\rm p})$ (nb),}   \\[3pt]
       & \multirowcell{1}{$N_{\jpsi}^{\rm diss}$} & \multirowcell{1}{$\mathrm{d}\sigma^{\rm diss}_{\jpsi}/\mathrm{d}y$ ({\textmu}b)} &  & & & \multirowcell{1}{$\sigma(\gamma+{\rm p} \rightarrow \jpsi+ {\rm p}^{(*)})$ (nb)}    \\[4pt]
    \hline
    \\[-10pt]
    \multirowcell{2}{(2.5, 4)} & $1180 \pm 84$ & $8.13 \pm 0.58 \pm 0.43$ & \multirowcell{2}{$209 \pm 4$} & \multirowcell{2}{(27, 57)} & \multirowcell{2}{39.9} & $39.0 \pm 2.8 \pm 2.2$ \\[2pt]
    & $1515 \pm 83$  & $10.43 \pm 0.57 \pm 1.39$ & & & & $50.0 \pm 2.7 \pm 6.7$ \\[2pt]
    \hline
    \\[-10pt]
    \multirowcell{2}{(3.25, 4)} & $564 \pm 53$  & $7.16 \pm 0.67 \pm 0.48$ & \multirowcell{2}{$220 \pm 4$} & \multirowcell{2}{(27, 39)} & \multirowcell{2}{32.8} & $32.51 \pm 3.0 \pm 2.3$ \\[2pt]
    & $733 \pm 52$ & $9.31 \pm 0.66 \pm 1.28$ & & & & $42.3 \pm 3.0 \pm 5.9$ \\[2pt]
    \multirowcell{2}{(2.5, 3.25)} & $629 \pm 54$ & $9.21 \pm 0.80 \pm 0.51$ & \multirowcell{2}{$197 \pm 4$} & \multirowcell{2}{(39, 57)} & \multirowcell{2}{47.7} & $46.8 \pm 4.1 \pm 2.8$ \\[2pt]
    & $768 \pm 55$ & $11.26 \pm 0.80 \pm 1.53$ & & & & $57.2 \pm 4.1 \pm 7.8$ \\[2pt]
\hline

\end{tabular}
\end{table}

\subsubsection{Exclusive \texorpdfstring{\jpsi}{jpsi} photoproduction}

\begin{figure}[tb]
    \begin{center}
\includegraphics[width=0.7\textwidth]{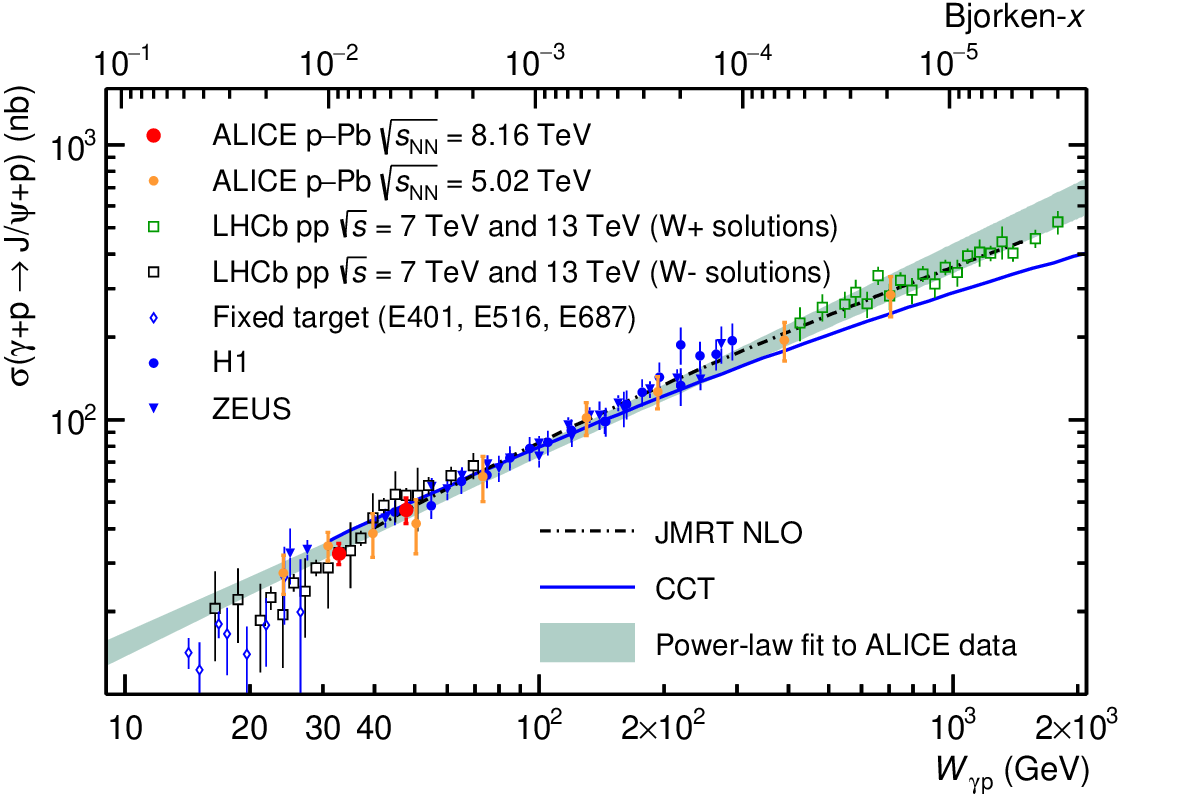}
\end{center}
\caption{Exclusive \jpsi photoproduction cross section off protons measured as a function of the centre-of-mass energy of the photon--proton system \Wgp by ALICE in \pPb UPCs and compared with previous measurements~\cite{Binkley:1981kv,Denby:1983wv,E687:1993aa,TheALICE:2014dwa, ALICE:2018oyo,H1:2005dtp, H1:2013okq, ZEUS:2002wfj,Aaij:2014iea, LHCb:2013nqs,LHCb:2018rcm} and with next-to-leading-order JMRT~\cite{Jones:2013pga, Jones:2016icr} and CCT~\cite{Cepila:2016uku} models. 
The power-law fit to the ALICE data is also shown. 
The uncertainties of the data points are the quadratic sum of the statistical and systematic uncertainties.
}
\label{fig:jpsi-crossX-exc}
\end{figure}

Figure~\ref{fig:jpsi-crossX-exc} shows the exclusive \jpsi photoproduction cross section $\sigma(\gamma + {\rm p } \rightarrow \jpsi + {\rm p})$  reported in Table~\ref{tab:crosssections-jpsi} 
 as a function of \Wgp, covering the range 27 $< \Wgp <$ 57~GeV. Comparisons with previous measurements and with several theoretical models are also shown.

Measurements at low \Wgp were performed by fixed target experiments, such as those reported by the E401~\cite{Binkley:1981kv}, E516~\cite{Denby:1983wv} and E687~\cite{E687:1993aa} Collaborations. Recently, measurements were performed near threshold by the GlueX Collaboration~\cite{GlueX:2019mkq} and by the E12-16-007 experiment~\cite{Duran:2022xag} which are not shown in Fig.~\ref{fig:jpsi-crossX-exc} since they fall outside of the power-law applicability discussed below. 

The cross sections are also compared with previous ALICE results in \pPb at \fivenn~\cite{TheALICE:2014dwa, ALICE:2018oyo}, at forward, mid, and backward rapidity, covering the energy range $21 < \Wgp < 952$~\GeV.

In this analysis, a $\chi^2$ fit of a power-law function,
$N(\Wgp/W_0)^\delta$,
is performed to the two ALICE data sets at \eightsixteen and \fivenn together, with $W_0 = 90.0$~GeV, as done in HERA analyses~\cite{H1:2005dtp, H1:2013okq, ZEUS:2002wfj} 
and for previous ALICE measurements~\cite{ALICE:2018oyo}.
The technique follows what was done by the H1 Collaboration~\cite{H1:2009jxj} and the fit takes into account the statistical and systematic uncertainties. 
The parameters obtained from the fit are $N = 71.6 \pm 3.7$ nb and $\delta = 0.70 \pm 0.04$ with a correlation of $+0.16$ between the two parameters. The quality of the fit is $\chi^2$/ndf $= 1.62$ for 9 degrees of freedom. 
The value of the exponent is the same as in previous ALICE measurements~\cite{ALICE:2018oyo}. 
The H1 and ZEUS measurements, performed over an energy range \Wgp that encompasses the new ALICE measurements, are also shown in the same figure. 
They, respectively, found $\delta = 0.69 \pm 0.02 \text{ (stat.)} \pm 0.03 \text{ (syst.)}$, and $\delta = 0.67 \pm 0.03$ (tot.)~\cite{H1:2005dtp, H1:2013okq, ZEUS:2002wfj}. 
Thus the measurements by ALICE are compatible with the values measured by HERA experiments, and no deviation from a power law is observed up to about 700~\GeV.

LHCb measured the exclusive \jpsi photoproduction cross sections in \pp collisions, at \snn $= 7$~\TeV~\cite{Aaij:2014iea, LHCb:2013nqs} and 13~\TeV~\cite{LHCb:2018rcm}. 
The LHCb analyses use data from a symmetric system and thus suffer from the ambiguity in identifying the photon emitter and the photon target.
Since the non-exclusive \jpsi photoproduction depends on \Wgp, these processes are difficult to subtract and make the extraction of the underlying $\sigma (\Wgp)$ strongly model dependent. 
Moreover, the uncertainty in the hadronic survival probability in \pp collisions is much larger than in \pPb collisions, and samples of \pp collisions can contain a contamination of \jpsi production through Odderon--Pomeron fusion~\cite{CDF:2009xey, 1997-eta}. 
For each $\mathrm{d} \sigma/\mathrm{d} y$ measurement, LHCb reported two solutions, one for low \Wgp and one for high \Wgp. 
Despite these ambiguities and assumptions, the LHCb solutions are found to be compatible with ALICE measurements within the current uncertainties.

ALICE measurements are also compared with the Jones--Martin--Ryskin--Teubner (JMRT) calculation. 
Two calculations are available from the JMRT group~\cite{Jones:2013pga, Jones:2016icr}. The first one, referred to as LO, is based on a power-law description of the process from the result in Ref.~\cite{Ryskin:1992ui}, while the second one, labeled as NLO, includes contributions which mimic effects expected from the dominant NLO corrections. At high \Wgp, they deviate from a simple power-law shape. 
Both models are fitted to the same data and their energy dependence is rather similar, so only the NLO version is shown.
ALICE measurements at \fivenn and \eightsixteen support their extracted gluon distribution down to $x \sim 2 \times 10^{-5}$. 
A more recent NLO computation of this process suggests a stronger sensitivity to quark contributions than previously considered~\cite{Eskola:2022vpi}.

Figure~\ref{fig:jpsi-crossX-exc} also shows predictions from the Cepila--Contreras--Takaki (CCT) model~\cite{Cepila:2016uku} based on the colour dipole approach. This model incorporates a fluctuating hot spot structure of the proton in the impact parameter plane, with the number of hot spots growing with decreasing $x$. It is compatible with ALICE measurements at \fivenn and \eightsixteen. Future UPC measurements by ALICE will explore the high $W$ range, particularly with future detector upgrades such as FoCal~\cite{Bylinkin:2022wkm}.

\subsubsection{Dissociative \texorpdfstring{\jpsi}{jpsi} photoproduction}

\begin{figure}[tb]
    \begin{center}
    \includegraphics[width=0.6\textwidth]{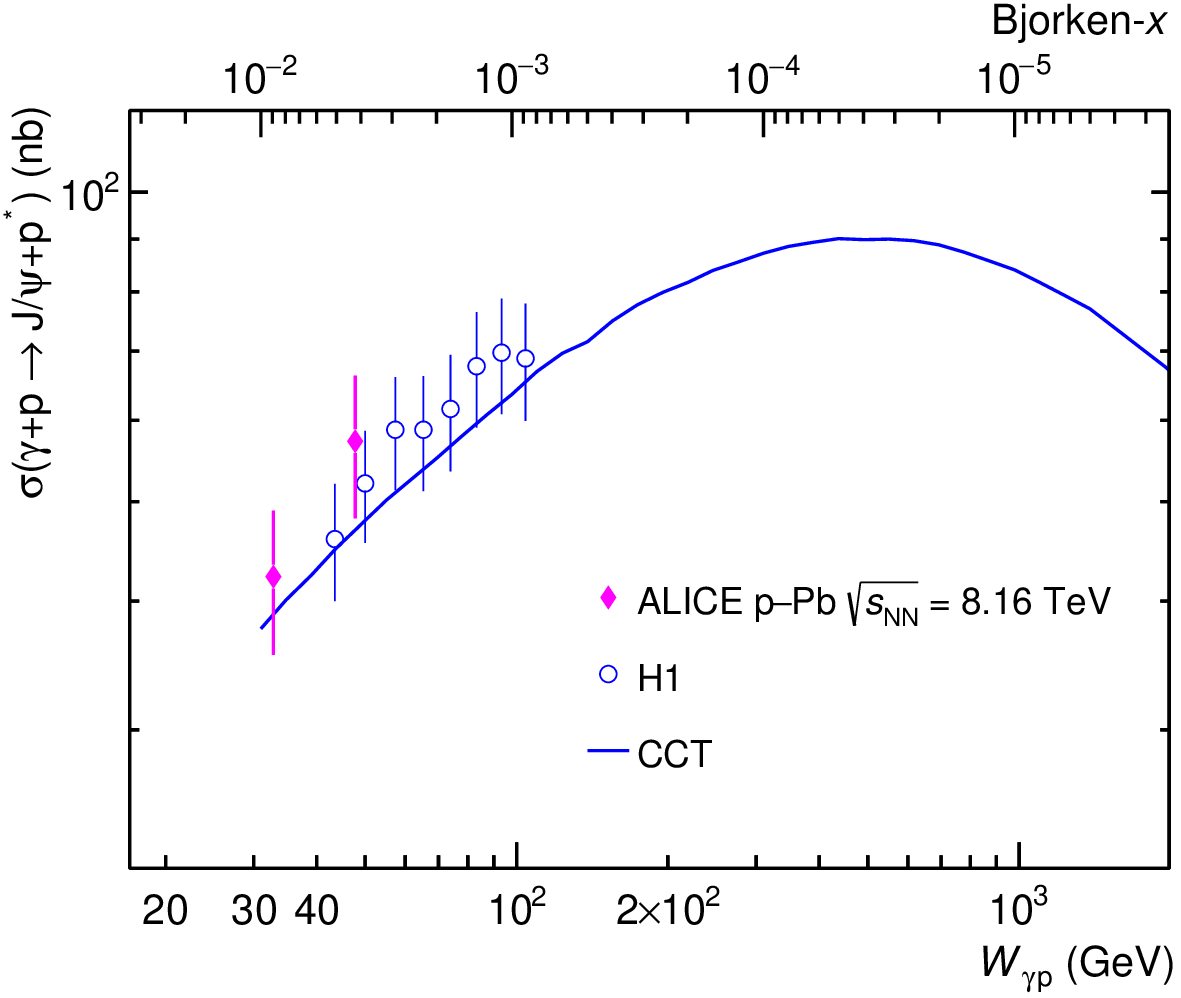}
    \end{center}
    \caption{Dissociative \jpsi photoproduction cross section off protons measured by ALICE in \pPb UPCs  at \eightsixteen and compared with H1 data~\cite{H1:2013okq}. A comparison with the CCT model~\cite{Cepila:2016uku} is shown. 
   The uncertainties of the data points are the quadratic sum of the statistical and
systematic uncertainties.
    }
\label{fig:jpsi-crossX-diss}
\end{figure}

Figure~\ref{fig:jpsi-crossX-diss} shows the ALICE measurement of the dissociative \jpsi photoproduction cross section $\sigma(\gamma + {\rm p } \rightarrow \jpsi + {\rm p}^{(*)})$ as a function of \Wgp, covering the range $27 < \Wgp < 57$~\GeV. 
The cross sections are also reported in Table~\ref{tab:crosssections-jpsi}. 
A previous measurement at similar energies by H1~\cite{H1:2013okq} is also shown and is in good agreement with the ALICE measurement.
In addition, the experimental results are compared with the CCT model~\cite{Cepila:2016uku} discussed in the previous section. 
In the framework of this model, the exclusive cross section is sensitive to the average interaction of the colour dipole ${\rm q}\overline{\rm q}$ with the proton, and the dissociative cross section is sensitive to the fluctuations in the ${\rm q}\overline{\rm q}$--proton interaction between the different colour field configurations of the proton. 
The model describes correctly the energy evolution of the dissociative cross section both for H1 and ALICE measurements and predicts that the cross section will reach a maximum at $\Wgp \simeq 500$~\GeV, then decrease at higher energies. This behaviour is expected due to the hot spots saturating the proton area.

\subsubsection{Ratio of dissociative-to-exclusive \texorpdfstring{\jpsi}{jpsi} photoproduction}

\begin{figure}[tb]
    \begin{center}
\includegraphics[width=0.6\textwidth]{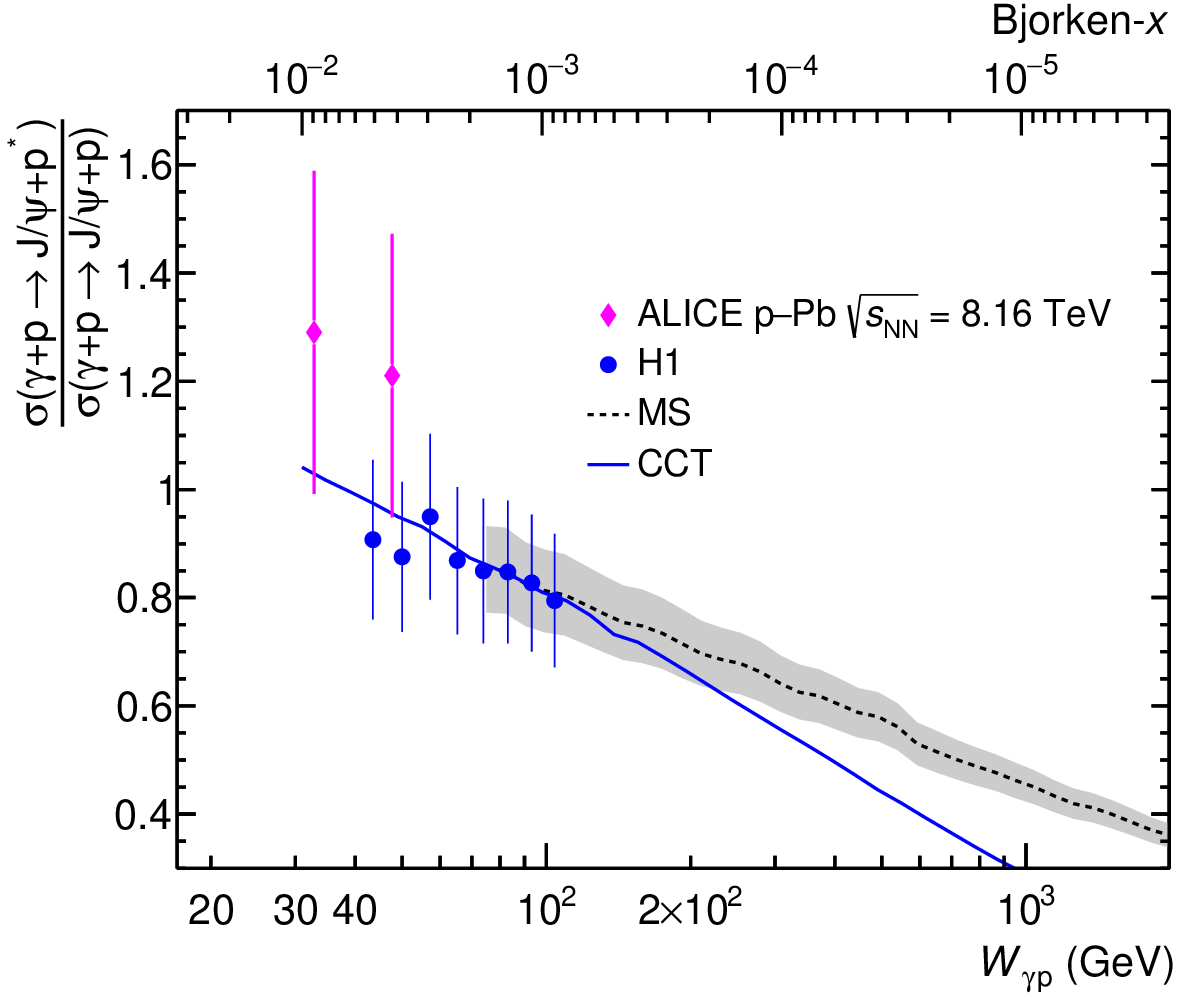}
\end{center}
\caption{
Ratio of dissociative-to-exclusive \jpsi photoproduction cross sections measured by ALICE in \pPb UPCs  at \eightsixteen  and compared with H1 measurements~\cite{H1:2013okq}. 
The uncertainties of the data points are the quadratic sum of the statistical and
systematic uncertainties.
The experimental uncertainties for the H1 data are computed assuming completely independent uncertainties for the exclusive and dissociative cross sections. 
The measurements are compared with the CCT model~\cite{Cepila:2016uku} and a model by Mäntysaari--Schenke (MS)~\cite{Mantysaari:2018zdd}. The uncertainty band of the MS model corresponds to the statistical uncertainty of the calculation. 
}
\label{fig:jpsi-crossX-ratio}
\end{figure}

ALICE measurements for the ratio of dissociative-to-exclusive \jpsi photoproduction cross sections, $\sigma(\gamma + {\rm p } \rightarrow \jpsi + {\rm p}^{(*)})/\sigma(\gamma + {\rm p } \rightarrow \jpsi + \rm{p})$, are given in Table~\ref{tab:ratio-jpsi}.
These measurements are also shown in Fig.~\ref{fig:jpsi-crossX-ratio} as a function of \Wgp, together with the measurements by H1~\cite{H1:2013okq} at similar energies. 
Two models are compared with the measurements: the CCT model~\cite{Cepila:2016uku}, and a model calculation by Mäntysaari--Schenke (MS)~\cite{Mantysaari:2018zdd}. The MS model is based on the perturbative JIMWLK (Jalilian-Iancu-McLerran-Weigert-Leonidov-Kovner) evolution~\cite{Jalilian-Marian:1996mkd,Jalilian-Marian:1997qno}, with initial parameters constrained from fits to H1 data starting from 
$x \sim 10^{-3}$. At high \Wgp, where the gluon saturation regime is expected, the models predict that the ratio of dissociative-to-exclusive cross sections vanishes.

\begin{table}[tb]
\centering
\caption{Ratio of dissociative-to-exclusive \jpsi photoproduction cross sections in \pPb UPCs at \eightsixteen. 
The first uncertainty is the statistical one. Its size is strongly impacted by the anti-correlation between exclusive and dissociative \jpsi components in the two-dimensional fit. The second uncertainty is the systematic one. It is computed as the quadratic sum of the signal extraction ratio uncertainty, 
and the uncertainty on the \VZEROC veto.}
\label{tab:ratio-jpsi}
\small
\begin{tabular}{lllllll}
\hline
    \multirowcell{2}{ Rapidity range }   & \multirowcell{2}{$\Wgp$ \\ (GeV)}    &  \multirowcell{2}{$\langle \Wgp \rangle$ \\ (GeV)} & \multirowcell{2}{$\displaystyle \frac{\sigma(\gamma+ {\rm p} \rightarrow \jpsi+ {\rm p}^{(*)})}{\sigma(\gamma+ {\rm p} \rightarrow \jpsi+ {\rm p})}$ }   \\
     & & \\
     & & \\[-10pt]
     \hline
    $(2.5, 4)$ & $(27, 57)$ & $39.9$ & $1.27 \pm 0.15 \pm 0.18$  \\
    \hline
    $(3.25, 4)$ & $(27, 39)$ & $32.8$ & $1.29 \pm 0.23 \pm 0.19$ \\
    $(2.5, 3.25)$ & $(39, 57)$ & $47.7$ & $1.21 \pm 0.18 \pm 0.18$ \\
\hline

\end{tabular}
\end{table}

\section{Summary}
This article presents three different measurements carried out by the ALICE Collaboration  in ultra-peripheral \pPb collisions at \eightsixteen. The exclusive dimuon continuum production from two-photon interactions in the invariant mass range from 1 to 2.5~\GeVmass is presented. It is compared with \starlight and \superchic and found to be compatible within three standard deviations. Since these models are based on LO QED calculations, this measurement can be used to provide a limit on higher-order corrections for this process.  
Furthermore, the  exclusive and dissociative  \jpsi photoproductions  off protons were measured. The measurement of exclusive \jpsi photoproduction cross section  is compared with those previously performed by ALICE, LHCb, H1, and ZEUS Collaborations. The ALICE measurements are consistent with a power-law dependence 
on \Wgp of $\sigma(\gp \to \jpsi {\rm p})$,
with the power found to be $\delta = 0.70 \pm 0.04$. 
The measurement of the cross section of dissociative photoproduction of \jpsi mesons is the first of its kind at the LHC and a first measurement of this type at a hadron collider. It is in good agreement with H1 measurements.
 This is the first step to probe the fluctuation of the subnucleonic structure in protons in ultra-peripheral collisions at high energies.


\newenvironment{acknowledgement}{\relax}{\relax}
\begin{acknowledgement}
\section*{Acknowledgements}

The ALICE Collaboration would like to thank all its engineers and technicians for their invaluable contributions to the construction of the experiment and the CERN accelerator teams for the outstanding performance of the LHC complex.
The ALICE Collaboration gratefully acknowledges the resources and support provided by all Grid centres and the Worldwide LHC Computing Grid (WLCG) collaboration.
The ALICE Collaboration acknowledges the following funding agencies for their support in building and running the ALICE detector:
A. I. Alikhanyan National Science Laboratory (Yerevan Physics Institute) Foundation (ANSL), State Committee of Science and World Federation of Scientists (WFS), Armenia;
Austrian Academy of Sciences, Austrian Science Fund (FWF): [M 2467-N36] and Nationalstiftung f\"{u}r Forschung, Technologie und Entwicklung, Austria;
Ministry of Communications and High Technologies, National Nuclear Research Center, Azerbaijan;
Conselho Nacional de Desenvolvimento Cient\'{\i}fico e Tecnol\'{o}gico (CNPq), Financiadora de Estudos e Projetos (Finep), Funda\c{c}\~{a}o de Amparo \`{a} Pesquisa do Estado de S\~{a}o Paulo (FAPESP) and Universidade Federal do Rio Grande do Sul (UFRGS), Brazil;
Bulgarian Ministry of Education and Science, within the National Roadmap for Research Infrastructures 2020-2027 (object CERN), Bulgaria;
Ministry of Education of China (MOEC) , Ministry of Science \& Technology of China (MSTC) and National Natural Science Foundation of China (NSFC), China;
Ministry of Science and Education and Croatian Science Foundation, Croatia;
Centro de Aplicaciones Tecnol\'{o}gicas y Desarrollo Nuclear (CEADEN), Cubaenerg\'{\i}a, Cuba;
Ministry of Education, Youth and Sports of the Czech Republic, Czech Republic;
The Danish Council for Independent Research | Natural Sciences, the VILLUM FONDEN and Danish National Research Foundation (DNRF), Denmark;
Helsinki Institute of Physics (HIP), Finland;
Commissariat \`{a} l'Energie Atomique (CEA) and Institut National de Physique Nucl\'{e}aire et de Physique des Particules (IN2P3) and Centre National de la Recherche Scientifique (CNRS), France;
Bundesministerium f\"{u}r Bildung und Forschung (BMBF) and GSI Helmholtzzentrum f\"{u}r Schwerionenforschung GmbH, Germany;
General Secretariat for Research and Technology, Ministry of Education, Research and Religions, Greece;
National Research, Development and Innovation Office, Hungary;
Department of Atomic Energy Government of India (DAE), Department of Science and Technology, Government of India (DST), University Grants Commission, Government of India (UGC) and Council of Scientific and Industrial Research (CSIR), India;
National Research and Innovation Agency - BRIN, Indonesia;
Istituto Nazionale di Fisica Nucleare (INFN), Italy;
Japanese Ministry of Education, Culture, Sports, Science and Technology (MEXT) and Japan Society for the Promotion of Science (JSPS) KAKENHI, Japan;
Consejo Nacional de Ciencia (CONACYT) y Tecnolog\'{i}a, through Fondo de Cooperaci\'{o}n Internacional en Ciencia y Tecnolog\'{i}a (FONCICYT) and Direcci\'{o}n General de Asuntos del Personal Academico (DGAPA), Mexico;
Nederlandse Organisatie voor Wetenschappelijk Onderzoek (NWO), Netherlands;
The Research Council of Norway, Norway;
Commission on Science and Technology for Sustainable Development in the South (COMSATS), Pakistan;
Pontificia Universidad Cat\'{o}lica del Per\'{u}, Peru;
Ministry of Education and Science, National Science Centre and WUT ID-UB, Poland;
Korea Institute of Science and Technology Information and National Research Foundation of Korea (NRF), Republic of Korea;
Ministry of Education and Scientific Research, Institute of Atomic Physics, Ministry of Research and Innovation and Institute of Atomic Physics and University Politehnica of Bucharest, Romania;
Ministry of Education, Science, Research and Sport of the Slovak Republic, Slovakia;
National Research Foundation of South Africa, South Africa;
Swedish Research Council (VR) and Knut \& Alice Wallenberg Foundation (KAW), Sweden;
European Organization for Nuclear Research, Switzerland;
Suranaree University of Technology (SUT), National Science and Technology Development Agency (NSTDA), Thailand Science Research and Innovation (TSRI) and National Science, Research and Innovation Fund (NSRF), Thailand;
Turkish Energy, Nuclear and Mineral Research Agency (TENMAK), Turkey;
National Academy of  Sciences of Ukraine, Ukraine;
Science and Technology Facilities Council (STFC), United Kingdom;
National Science Foundation of the United States of America (NSF) and United States Department of Energy, Office of Nuclear Physics (DOE NP), United States of America.
In addition, individual groups or members have received support from:
European Research Council, Strong 2020 - Horizon 2020 (grant nos. 950692, 824093), European Union;
Academy of Finland (Center of Excellence in Quark Matter) (grant nos. 346327, 346328), Finland;
Programa de Apoyos para la Superaci\'{o}n del Personal Acad\'{e}mico, UNAM, Mexico.

\end{acknowledgement}

\bibliographystyle{utphys}   
\bibliography{bibliography}

\newpage
\appendix

%
%

\section{The ALICE Collaboration}
\label{app:collab}
\begin{flushleft} 
\small

S.~Acharya\,\orcidlink{0000-0002-9213-5329}\,$^{\rm 125}$, 
D.~Adamov\'{a}\,\orcidlink{0000-0002-0504-7428}\,$^{\rm 86}$, 
A.~Adler$^{\rm 69}$, 
G.~Aglieri Rinella\,\orcidlink{0000-0002-9611-3696}\,$^{\rm 32}$, 
M.~Agnello\,\orcidlink{0000-0002-0760-5075}\,$^{\rm 29}$, 
N.~Agrawal\,\orcidlink{0000-0003-0348-9836}\,$^{\rm 50}$, 
Z.~Ahammed\,\orcidlink{0000-0001-5241-7412}\,$^{\rm 132}$, 
S.~Ahmad\,\orcidlink{0000-0003-0497-5705}\,$^{\rm 15}$, 
S.U.~Ahn\,\orcidlink{0000-0001-8847-489X}\,$^{\rm 70}$, 
I.~Ahuja\,\orcidlink{0000-0002-4417-1392}\,$^{\rm 37}$
% ...

\section*{Affiliation Notes}

$^{\rm I}$ Deceased\\
$^{\rm II}$ Also at: Max-Planck-Institut f\"{u}r Physik, Munich, Germany\\
$^{\rm III}$ Also at: Italian National Agency for New Technologies, Energy and Sustainable Economic Development (ENEA), Bologna, Italy\\
$^{\rm IV}$ Also at: Department of Applied Physics, Aligarh Muslim University, Aligarh, India\\
$^{\rm V}$ Also at: Institute of Theoretical Physics, University of Wroclaw, Poland\\
$^{\rm VI}$ Also at: An institution covered by a cooperation agreement with CERN\\

\section*{Collaboration Institutes}

$^{1}$ A.I. Alikhanyan National Science Laboratory (Yerevan Physics Institute) Foundation, Yerevan, Armenia\\
$^{2}$ AGH University of Science and Technology, Cracow, Poland\\
$^{3}$ Bogolyubov Institute for Theoretical Physics, National Academy of Sciences of Ukraine, Kiev, Ukraine\\
$^{4}$ Bose Institute, Department of Physics  and Centre for Astroparticle Physics and Space Science (CAPSS), Kolkata, India\\
$^{5}$ California Polytechnic State University, San Luis Obispo, California, United States\\
$^{6}$ Central China Normal University, Wuhan, China\\
$^{7}$ Centro de Aplicaciones Tecnol\'{o}gicas y Desarrollo Nuclear (CEADEN), Havana, Cuba\\
$^{8}$ Centro de Investigaci\'{o}n y de Estudios Avanzados (CINVESTAV), Mexico City and M\'{e}rida, Mexico\\
$^{9}$ Chicago State University, Chicago, Illinois, United States\\
$^{10}$ China Institute of Atomic Energy, Beijing, China\\
$^{11}$ Chungbuk National University, Cheongju, Republic of Korea\\
$^{12}$ Comenius University Bratislava, Faculty of Mathematics, Physics and Informatics, Bratislava, Slovak Republic\\
$^{13}$ COMSATS University Islamabad, Islamabad, Pakistan\\
$^{14}$ Creighton University, Omaha, Nebraska, United States\\
$^{15}$ Department of Physics, Aligarh Muslim University, Aligarh, India\\
$^{16}$ Department of Physics, Pusan National University, Pusan, Republic of Korea\\
$^{17}$ Department of Physics, Sejong University, Seoul, Republic of Korea\\
$^{18}$ Department of Physics, University of California, Berkeley, California, United States\\
$^{19}$ Department of Physics, University of Oslo, Oslo, Norway\\
$^{20}$ Department of Physics and Technology, University of Bergen, Bergen, Norway\\
$^{21}$ Dipartimento di Fisica, Universit\`{a} di Pavia, Pavia, Italy\\
$^{22}$ Dipartimento di Fisica dell'Universit\`{a} and Sezione INFN, Cagliari, Italy\\
$^{23}$ Dipartimento di Fisica dell'Universit\`{a} and Sezione INFN, Trieste, Italy\\
$^{24}$ Dipartimento di Fisica dell'Universit\`{a} and Sezione INFN, Turin, Italy\\
$^{25}$ Dipartimento di Fisica e Astronomia dell'Universit\`{a} and Sezione INFN, Bologna, Italy\\
$^{26}$ Dipartimento di Fisica e Astronomia dell'Universit\`{a} and Sezione INFN, Catania, Italy\\
$^{27}$ Dipartimento di Fisica e Astronomia dell'Universit\`{a} and Sezione INFN, Padova, Italy\\
$^{28}$ Dipartimento di Fisica `E.R.~Caianiello' dell'Universit\`{a} and Gruppo Collegato INFN, Salerno, Italy\\
$^{29}$ Dipartimento DISAT del Politecnico and Sezione INFN, Turin, Italy\\
$^{30}$ Dipartimento di Scienze MIFT, Universit\`{a} di Messina, Messina, Italy\\
$^{31}$ Dipartimento Interateneo di Fisica `M.~Merlin' and Sezione INFN, Bari, Italy\\
$^{32}$ European Organization for Nuclear Research (CERN), Geneva, Switzerland\\
$^{33}$ Faculty of Electrical Engineering, Mechanical Engineering and Naval Architecture, University of Split, Split, Croatia\\
$^{34}$ Faculty of Engineering and Science, Western Norway University of Applied Sciences, Bergen, Norway\\
$^{35}$ Faculty of Nuclear Sciences and Physical Engineering, Czech Technical University in Prague, Prague, Czech Republic\\
$^{36}$ Faculty of Physics, Sofia University, Sofia, Bulgaria\\
$^{37}$ Faculty of Science, P.J.~\v{S}af\'{a}rik University, Ko\v{s}ice, Slovak Republic\\
$^{38}$ Frankfurt Institute for Advanced Studies, Johann Wolfgang Goethe-Universit\"{a}t Frankfurt, Frankfurt, Germany\\
$^{39}$ Fudan University, Shanghai, China\\
$^{40}$ Gangneung-Wonju National University, Gangneung, Republic of Korea\\
$^{41}$ Gauhati University, Department of Physics, Guwahati, India\\
$^{42}$ Helmholtz-Institut f\"{u}r Strahlen- und Kernphysik, Rheinische Friedrich-Wilhelms-Universit\"{a}t Bonn, Bonn, Germany\\
$^{43}$ Helsinki Institute of Physics (HIP), Helsinki, Finland\\
$^{44}$ High Energy Physics Group,  Universidad Aut\'{o}noma de Puebla, Puebla, Mexico\\
$^{45}$ Horia Hulubei National Institute of Physics and Nuclear Engineering, Bucharest, Romania\\
$^{46}$ Indian Institute of Technology Bombay (IIT), Mumbai, India\\
$^{47}$ Indian Institute of Technology Indore, Indore, India\\
$^{48}$ INFN, Laboratori Nazionali di Frascati, Frascati, Italy\\
$^{49}$ INFN, Sezione di Bari, Bari, Italy\\
$^{50}$ INFN, Sezione di Bologna, Bologna, Italy\\
$^{51}$ INFN, Sezione di Cagliari, Cagliari, Italy\\
$^{52}$ INFN, Sezione di Catania, Catania, Italy\\
$^{53}$ INFN, Sezione di Padova, Padova, Italy\\
$^{54}$ INFN, Sezione di Pavia, Pavia, Italy\\
$^{55}$ INFN, Sezione di Torino, Turin, Italy\\
$^{56}$ INFN, Sezione di Trieste, Trieste, Italy\\
$^{57}$ Inha University, Incheon, Republic of Korea\\
$^{58}$ Institute for Gravitational and Subatomic Physics (GRASP), Utrecht University/Nikhef, Utrecht, Netherlands\\
$^{59}$ Institute of Experimental Physics, Slovak Academy of Sciences, Ko\v{s}ice, Slovak Republic\\
$^{60}$ Institute of Physics, Homi Bhabha National Institute, Bhubaneswar, India\\
$^{61}$ Institute of Physics of the Czech Academy of Sciences, Prague, Czech Republic\\
$^{62}$ Institute of Space Science (ISS), Bucharest, Romania\\
$^{63}$ Institut f\"{u}r Kernphysik, Johann Wolfgang Goethe-Universit\"{a}t Frankfurt, Frankfurt, Germany\\
$^{64}$ Instituto de Ciencias Nucleares, Universidad Nacional Aut\'{o}noma de M\'{e}xico, Mexico City, Mexico\\
$^{65}$ Instituto de F\'{i}sica, Universidade Federal do Rio Grande do Sul (UFRGS), Porto Alegre, Brazil\\
$^{66}$ Instituto de F\'{\i}sica, Universidad Nacional Aut\'{o}noma de M\'{e}xico, Mexico City, Mexico\\
$^{67}$ iThemba LABS, National Research Foundation, Somerset West, South Africa\\
$^{68}$ Jeonbuk National University, Jeonju, Republic of Korea\\
$^{69}$ Johann-Wolfgang-Goethe Universit\"{a}t Frankfurt Institut f\"{u}r Informatik, Fachbereich Informatik und Mathematik, Frankfurt, Germany\\
$^{70}$ Korea Institute of Science and Technology Information, Daejeon, Republic of Korea\\
$^{71}$ KTO Karatay University, Konya, Turkey\\
$^{72}$ Laboratoire de Physique des 2 Infinis, Ir\`{e}ne Joliot-Curie, Orsay, France\\
$^{73}$ Laboratoire de Physique Subatomique et de Cosmologie, Universit\'{e} Grenoble-Alpes, CNRS-IN2P3, Grenoble, France\\
$^{74}$ Lawrence Berkeley National Laboratory, Berkeley, California, United States\\
$^{75}$ Lund University Department of Physics, Division of Particle Physics, Lund, Sweden\\
$^{76}$ Nagasaki Institute of Applied Science, Nagasaki, Japan\\
$^{77}$ Nara Women{'}s University (NWU), Nara, Japan\\
$^{78}$ National and Kapodistrian University of Athens, School of Science, Department of Physics , Athens, Greece\\
$^{79}$ National Centre for Nuclear Research, Warsaw, Poland\\
$^{80}$ National Institute of Science Education and Research, Homi Bhabha National Institute, Jatni, India\\
$^{81}$ National Nuclear Research Center, Baku, Azerbaijan\\
$^{82}$ National Research and Innovation Agency - BRIN, Jakarta, Indonesia\\
$^{83}$ Niels Bohr Institute, University of Copenhagen, Copenhagen, Denmark\\
$^{84}$ Nikhef, National institute for subatomic physics, Amsterdam, Netherlands\\
$^{85}$ Nuclear Physics Group, STFC Daresbury Laboratory, Daresbury, United Kingdom\\
$^{86}$ Nuclear Physics Institute of the Czech Academy of Sciences, Husinec-\v{R}e\v{z}, Czech Republic\\
$^{87}$ Oak Ridge National Laboratory, Oak Ridge, Tennessee, United States\\
$^{88}$ Ohio State University, Columbus, Ohio, United States\\
$^{89}$ Physics department, Faculty of science, University of Zagreb, Zagreb, Croatia\\
$^{90}$ Physics Department, Panjab University, Chandigarh, India\\
$^{91}$ Physics Department, University of Jammu, Jammu, India\\
$^{92}$ Physics Program and International Institute for Sustainability with Knotted Chiral Meta Matter (SKCM2), Hiroshima University, Hiroshima, Japan\\
$^{93}$ Physikalisches Institut, Eberhard-Karls-Universit\"{a}t T\"{u}bingen, T\"{u}bingen, Germany\\
$^{94}$ Physikalisches Institut, Ruprecht-Karls-Universit\"{a}t Heidelberg, Heidelberg, Germany\\
$^{95}$ Physik Department, Technische Universit\"{a}t M\"{u}nchen, Munich, Germany\\
$^{96}$ Politecnico di Bari and Sezione INFN, Bari, Italy\\
$^{97}$ Research Division and ExtreMe Matter Institute EMMI, GSI Helmholtzzentrum f\"ur Schwerionenforschung GmbH, Darmstadt, Germany\\
$^{98}$ Saga University, Saga, Japan\\
$^{99}$ Saha Institute of Nuclear Physics, Homi Bhabha National Institute, Kolkata, India\\
$^{100}$ School of Physics and Astronomy, University of Birmingham, Birmingham, United Kingdom\\
$^{101}$ Secci\'{o}n F\'{\i}sica, Departamento de Ciencias, Pontificia Universidad Cat\'{o}lica del Per\'{u}, Lima, Peru\\
$^{102}$ Stefan Meyer Institut f\"{u}r Subatomare Physik (SMI), Vienna, Austria\\
$^{103}$ SUBATECH, IMT Atlantique, Nantes Universit\'{e}, CNRS-IN2P3, Nantes, France\\
$^{104}$ Sungkyunkwan University, Suwon City, Republic of Korea\\
$^{105}$ Suranaree University of Technology, Nakhon Ratchasima, Thailand\\
$^{106}$ Technical University of Ko\v{s}ice, Ko\v{s}ice, Slovak Republic\\
$^{107}$ The Henryk Niewodniczanski Institute of Nuclear Physics, Polish Academy of Sciences, Cracow, Poland\\
$^{108}$ The University of Texas at Austin, Austin, Texas, United States\\
$^{109}$ Universidad Aut\'{o}noma de Sinaloa, Culiac\'{a}n, Mexico\\
$^{110}$ Universidade de S\~{a}o Paulo (USP), S\~{a}o Paulo, Brazil\\
$^{111}$ Universidade Estadual de Campinas (UNICAMP), Campinas, Brazil\\
$^{112}$ Universidade Federal do ABC, Santo Andre, Brazil\\
$^{113}$ University of Cape Town, Cape Town, South Africa\\
$^{114}$ University of Houston, Houston, Texas, United States\\
$^{115}$ University of Jyv\"{a}skyl\"{a}, Jyv\"{a}skyl\"{a}, Finland\\
$^{116}$ University of Kansas, Lawrence, Kansas, United States\\
$^{117}$ University of Liverpool, Liverpool, United Kingdom\\
$^{118}$ University of Science and Technology of China, Hefei, China\\
$^{119}$ University of South-Eastern Norway, Kongsberg, Norway\\
$^{120}$ University of Tennessee, Knoxville, Tennessee, United States\\
$^{121}$ University of the Witwatersrand, Johannesburg, South Africa\\
$^{122}$ University of Tokyo, Tokyo, Japan\\
$^{123}$ University of Tsukuba, Tsukuba, Japan\\
$^{124}$ University Politehnica of Bucharest, Bucharest, Romania\\
$^{125}$ Universit\'{e} Clermont Auvergne, CNRS/IN2P3, LPC, Clermont-Ferrand, France\\
$^{126}$ Universit\'{e} de Lyon, CNRS/IN2P3, Institut de Physique des 2 Infinis de Lyon, Lyon, France\\
$^{127}$ Universit\'{e} de Strasbourg, CNRS, IPHC UMR 7178, F-67000 Strasbourg, France, Strasbourg, France\\
$^{128}$ Universit\'{e} Paris-Saclay Centre d'Etudes de Saclay (CEA), IRFU, D\'{e}partment de Physique Nucl\'{e}aire (DPhN), Saclay, France\\
$^{129}$ Universit\`{a} degli Studi di Foggia, Foggia, Italy\\
$^{130}$ Universit\`{a} del Piemonte Orientale, Vercelli, Italy\\
$^{131}$ Universit\`{a} di Brescia, Brescia, Italy\\
$^{132}$ Variable Energy Cyclotron Centre, Homi Bhabha National Institute, Kolkata, India\\
$^{133}$ Warsaw University of Technology, Warsaw, Poland\\
$^{134}$ Wayne State University, Detroit, Michigan, United States\\
$^{135}$ Westf\"{a}lische Wilhelms-Universit\"{a}t M\"{u}nster, Institut f\"{u}r Kernphysik, M\"{u}nster, Germany\\
$^{136}$ Wigner Research Centre for Physics, Budapest, Hungary\\
$^{137}$ Yale University, New Haven, Connecticut, United States\\
$^{138}$ Yonsei University, Seoul, Republic of Korea\\
$^{139}$  Zentrum  f\"{u}r Technologie und Transfer (ZTT), Worms, Germany\\
$^{140}$ Affiliated with an institute covered by a cooperation agreement with CERN\\
$^{141}$ Affiliated with an international laboratory covered by a cooperation agreement with CERN.\\

\end{flushleft} 

\end{document}